\documentclass[journal]{IEEEtran}
\usepackage{cite}
\ifCLASSINFOpdf
  \usepackage[pdftex]{graphicx}
  \graphicspath{{../pdf/}{../jpeg/}}
  \DeclareGraphicsExtensions{.pdf,.jpeg,.png}
\else
  \usepackage[dvips]{graphicx}
  \graphicspath{{../eps/}}
  \DeclareGraphicsExtensions{.eps}
\fi
\usepackage{amsmath}
\usepackage{amsfonts}
\usepackage[dvipsnames]{xcolor}
\usepackage{amssymb}
\usepackage{array}
\usepackage{multirow}
\usepackage{longtable}
\usepackage[english]{babel}
\usepackage{amsthm}
\usepackage{scalerel}
\usepackage{hyperref}

\usepackage{textcomp}
\usepackage{url}
\usepackage{xurl}
\usepackage{threeparttable}
\usepackage[
linesnumbered,ruled,vlined]{algorithm2e}
\usepackage {algpseudocode}
\usepackage{algorithmicx}
\usepackage{algcompatible}
\input{mysymbol.sty}

\makeatletter
\newcommand{\thickhline}{%
    \noalign {\ifnum 0=`}\fi \hrule height 1pt
    \futurelet \reserved@a \@xhline
}
\newcolumntype{"}{@{\hskip\tabcolsep\vrule width 1pt\hskip\tabcolsep}}
\makeatother
\ifCLASSOPTIONcompsoc
  \usepackage[caption=false,font=normalsize,labelfont=sf,textfont=sf]{subfig}
\else
  \usepackage[caption=false,font=footnotesize]{subfig}
\fi
\ifCLASSOPTIONcaptionsoff
  \usepackage[nomarkers]{endfloat}
 \let\MYoriglatexcaption\caption
 \renewcommand{\caption}[2][\relax]{\MYoriglatexcaption[#2]{#2}}
\fi
\let\MYorigsubfloat\subfloat
\renewcommand{\subfloat}[2][\relax]{\MYorigsubfloat[]{#2}}
\begin{document}
\title{Grid-Aware Assessment of Forced Oscillations \\ from Spatially Distributed Large Loads}

\author{Min-Seung~Ko,~\IEEEmembership{Member,~IEEE,} and
Hao~Zhu,~\IEEEmembership{Senior~Member,~IEEE}

\thanks{This work has been supported by NSF grants 2130706 and 2150571 (\textit{Corresponding Author: Hao Zhu.})}
\thanks{M.-S. Ko is with the School of Electrical and Electronic Engineering, Yonsei University, Seoul, 03722, Korea. (e-mail: minseungko@yonsei.ac.kr)}
\thanks{H. Zhu is with the Chandra Family Department of ECE, The University of Texas at Austin, Austin, TX 78712, USA. (e-mail: haozhu@utexas.edu)}}

\maketitle
 
\begin{abstract}
This paper develops a unified analytical framework for assessing forced oscillations induced by spatially distributed large-load fluctuations to  determine their grid-aware fluctuation magnitudes. Emerging computational loads, such as hyperscale datacenters, present cyclic power variations with dynamic impacts depending on the waveform characteristics, forcing frequency, and source location. Our grid-aware framework represents these fluctuations using deterministic harmonic and probabilistic spectral models,  and analyzes the resulting bus-wise frequency deviations based on a generic grid frequency-response model. Deterministic, risk-aware, and band-based probabilistic severity measures are further developed to obtain admissible fluctuation limits under different assumptions on frequency and phase uncertainty. We also formulate a security-constrained (SC) optimization of  locational fluctuation magnitudes among existing large loads, accounting for their aggregated frequency deviation level. Case studies on the IEEE 39-bus and 14-generator Australian systems validate the proposed severity measures and admissible limit analysis. The SC optimization results further show that distributing a given amount of aggregate fluctuation toward dynamically less sensitive locations can substantially reduce system-wide oscillation severity.
\end{abstract}

\section{Introduction}\label{sec1}
\IEEEPARstart{T}{he} rapid integration of large loads presents new dynamic challenges to power systems. Load growth has long been incorporated into power system planning studies through transmission expansion, generation investment, and adequacy-oriented assessment \cite{roh2007market,spyrou2017benefits}. However, emerging large electronic loads (LELs), particularly hyperscale AI datacenters, raise additional dynamic and stability concerns not yet fully captured by conventional planning studies \cite{sun2022dynamic,al2020manipulation,chen2025electricity}. Large AI training workloads can exhibit synchronized computation--communication cycles and abrupt workload transitions, resulting in repetitive power swings and steep ramping changes \cite{ko2026wide,choukse2025power,jensen2026easyrider}. Other computational loads, such as crypto-mining facilities, also exhibit substantial time-varying power demand \cite{hajiaghapour2024investigating}. When such variations contain sustained or recurring spectral components, they can act as external forcing inputs and excite grid electromechanical modes over a large interconnection. As these large loads continue to grow in capacity and geographical concentration, it is necessary to develop new analytical models and computational tools to assess their potential to drive grid-wide forced oscillations (FO).

FOs are sustained grid responses driven by periodic external disturbances, especially when the forcing frequency approaches poorly damped system modes \cite{follum2017power}. Extensive studies have focused on locating FO sources from measurement data using dissipating-energy-flow, effective-impedance, and data-driven approaches \cite{maslennikov2017dissipating,chevalier2018using,jiang2024forced,liu2024data}, to support post-event diagnosis. Another line of work has examined system vulnerability to potential forcing inputs. NERC reports \cite{nerc2021recommended,nerc2010interconnection} highlight the influence of harmonic components and the strong dependence on forcing frequency/location, while \cite{alshuaibi2022forced} provides a vulnerability evaluation approach by scanning disturbance frequencies and locations. However, existing work largely considers sinusoidal forcing from a small number of sources (no more than 2), limiting their applicability to emerging large loads with heterogeneous characteristics.

The dynamic characterization and impact analysis for large datacenters have attracted increasing research interest. One line of research has focused on improving their modeling for grid dynamic studies by explicitly representing power-electronic interfaces, internal controls, power-delivery architectures, and workload-dependent demand variations \cite{sun2022dynamic,jimenez2025data,ross2026electromagnetic,zhao2026dynamic}. Beyond modeling, recent analytical and numerical studies have shown that AI- and computation-driven load variations can excite poorly damped grid modes and produce oscillatory responses that propagate beyond the point of interconnection \cite{valverde2025forced,ko2026wide}. Field measurements have also documented sustained oscillatory events associated with datacenters and other LELs, highlighting the practical relevance of load-induced oscillations \cite{mishra2025inception,fan2026replicating}.

Meanwhile, industry guidance has emphasized that emerging LELs with variable or cyclical power demands require the development of new FO-oriented interconnection studies for both regional and inter-area modes \cite{esig2026large, nerc2026assessment,nerc2026risk}. Recent research has started to address these concerns with quantitative screening and operating criteria. The simulation-based framework in \cite{biswas2025evaluating} evaluates LEL-induced oscillation risk over various forcing frequencies/locations, while a frequency-domain approach in \cite{hossain2026limiting} analyzes the allowable load fluctuations and screens candidate locations using turbine-generator torsional-fatigue constraints. Moreover, an analytical resonance-safety criterion is developed in \cite{chaudhary2026predispatch} to bound the size of an AI training cluster. These studies, albeit important, tend to address relatively specialized objectives.
Despite these advances, a unified system-level, grid-aware analytical framework is still needed to characterize the complex effects of spatially distributed forcing sources. In particular, the waveform-dependent harmonic excitation and uncertainty in source frequencies/phases are very important for analyzing the system-wide electromechanical oscillations. Moreover, it is still imperative to use these response characteristics to design new planning and operating metrics for better quantifying the FO severity and determining admissible fluctuation levels \cite{nerc2026assessment}.

Therefore, this paper proposes a unified analytical framework for assessing FOs induced by spatially distributed large-load fluctuations. Our grid-aware assessment represents practical non-sinusoidal and ramp-limited fluctuations through deterministic harmonic models, while capturing spatiotemporal uncertainty using probabilistic spectral analysis. Based on a generic grid-response representation, we develop both deterministic and probabilistic severity metrics for the system-wide frequency deviations, and use them to analyze the admissible fluctuation limits based on deterministic, risk-aware, and probabilistic criteria. We further formulate the security-constrained (SC) optimization that accounts for the aggregated frequency-deviation level among spatially distributed forcing sources. In short, the main contributions of this paper are three-fold:
\begin{itemize}
    \item Developing a unified analytical framework mapping spatially distributed large-load fluctuations to system-wide frequency deviations, with heterogeneous, non-sinusoidal waveforms, ramp-rate constraints, and uncertainty.
    \item Developing deterministic and probabilistic severity metrics to accurately characterize grid-aware FO vulnerability across different load-fluctuation models and uncertainty conditions.
    \item Developing a security-constrained framework that can optimize locational fluctuation magnitudes among existing large loads, thereby reducing the aggregated oscillation severity relative to a naive uniform allocation.
\end{itemize}

The remainder of this paper is organized as follows. Sec.~\ref{sec2} presents the grid frequency response along with the deterministic and probabilistic load fluctuation models. Sec.~\ref{sec3} develops the FO severity metrics, followed by the corresponding admissible fluctuation limits in Sec.~\ref{sec4}. Sec.~\ref{sec5} presents the SC optimization framework. Sec.~\ref{sec6} provides the case studies, and Sec.~\ref{sec7} concludes the paper.

\section{System Modeling}\label{sec2}

This section establishes the grid frequency-response model for analyzing the impact of spatially distributed load fluctuations. First, we derive a small-signal dynamic model to represent the network-wide frequency response to distributed fluctuations. Second, we present the modeling of load fluctuations, including deterministic non-sinusoidal waveforms and probabilistic spectral representations, to capture the distinct transient behaviors of large loads.

\subsection{Grid Frequency Response Model}\label{sec2-1}

We consider a power system with $n_b$ buses, including $n_g$ generator buses and $n_l$ load buses. In general, its dynamics can be described by differential-algebraic equations (DAEs) with dynamic states in vector $\mathbf{x}$ and algebraic variables in $\mathbf{z}$, as follows:
\begin{equation}
    \begin{cases}
        \dot{\mathbf{x}} &= f(\mathbf{x},\mathbf{z},\mathbf{u}) \\
        \mathbf 0 &= g(\mathbf{x},\mathbf{z},\mathbf{u})
    \end{cases}
    \label{dae}
\end{equation}
where $\mathbf{u}=\left[ u_1 \cdots u_{n_l}\right]^{\top}$ collects all large load fluctuations, with $u_{\ell}(t)$ denoting  the input fluctuation per load bus $\ell$. 
The algebraic equation $g(\cdot)$ represents the network power-flow (PF) relations between nodal power injections and voltage phase angles through the transmission system. Linearizing it around an equilibrium point and eliminating the algebraic variables yields a linear input-output representation of the grid dynamics. We specialize this representation to a reduced electromechanical frequency-response model under the DC power flow (PF) approximation, yielding $\Delta\mathbf{P}=\mathbf{J}\Delta\boldsymbol{\theta}$ with the susceptance-weighted network Laplacian $\mathbf{J}$. Partitioning the buses into generator buses $\ccalG$ and load buses $\ccalL$, one can extract the frequency dynamics retained at $\ccalG$ by eliminating $\ccalL$ via Kron reduction. This yields a reduced coupling matrix among generator buses only, as given by
\begin{equation}
    \mathbf{L}_\mathrm{red}=\mathbf{J}_{\ccalG\ccalG}-\mathbf{J}_{\ccalG\ccalL}\mathbf{J}_{\ccalL\ccalL}^{-1}\mathbf{J}_{\ccalL\ccalG},~\mathrm{with}~~\mathbf{J}=\begin{bmatrix}
        \mathbf{J}_{\ccalG\ccalG} & \mathbf{J}_{\ccalG\ccalL} \\
        \mathbf{J}_{\ccalL\ccalG} & \mathbf{J}_{\ccalL\ccalL}
    \end{bmatrix}.
    \label{redlap}
\end{equation}

We next define the reduced state vector $\mathbf{x}=\left[\Delta\boldsymbol{\theta} \; \Delta\boldsymbol{\omega} \right]^{\top}\in\mathbb{R}^{2n_g}$ consisting of the generator rotor angle and speed deviations, respectively. Using the reduced Laplacian $ \mathbf{L}_\mathrm{red}$, one can represent the second-order dynamics by
\begin{align}
    \dot{\mathbf{x}}&=\underbrace{\begin{bmatrix}
    0 & \mathbf{I} \\
    -\mathbf{M}^{-1}\mathbf{L}_\mathrm{red} & -\mathbf{M}^{-1}\mathbf{D}
    \end{bmatrix}}_{\mathbf{A}} \mathbf{x}
    \underbrace{-\begin{bmatrix}
        0 \\ \mathbf{M}^{-1}
    \end{bmatrix} \mathbf{J}_{\ccalG\ccalL}\mathbf{J}_{\ccalL\ccalL}^{-1}}_{\mathbf{B}}\mathbf{u}. \label{ltisys1} \\
    \mathbf{y}& = \begin{bmatrix}
        \Delta\mathbf{f}_{\ccalG} \\ \Delta\mathbf{f}_{\ccalL}
    \end{bmatrix} = \underbrace{\frac{1}{2\pi}\begin{bmatrix}
    0 & \mathbf{I} \\ 0 & -\mathbf{J}_{\ccalL\ccalL}^{-1}\mathbf{J}_{\ccalL\ccalG}
    \end{bmatrix}}_{\mathbf{C}} \mathbf{x}. \label{ltisys2}
\end{align} 
Here, $\mathbf{M}$ and $\mathbf{D}$ denote the equivalent aggregate inertia and damping, incorporating the contributions from both SGs and IBR frequency controls; see \cite{poolla2017optimal} for further discussion. Active-power fluctuations $\mathbf{u}$ propagate through the network and appear as equivalent power imbalances in the generator buses through the matrix $\mathbf{B}$. 
The load-bus frequency deviation $\Delta\mathbf{f}_{\ccalL}$ is reconstructed from $\Delta\boldsymbol{\omega}$ through $-\mathbf{J}_{\ccalL\ccalL}^{-1}\mathbf{J}_{\ccalL\ccalG}$. Specifically, although differentiating the algebraic load-bus angle relation also introduces a term proportional to $\dot{\mathbf{u}}$, we omit this term to focus on the electromechanical frequency component.

Given the linearized system in \eqref{ltisys1}--\eqref{ltisys2}, the relationship between $\mathbf{u}$ and $\mathbf{y}$ can be characterized in the frequency domain through a transfer function matrix $\mathbf{H}(s)=\mathbf{C}(s\mathbf{I}-\mathbf{A})^{-1}\mathbf{B}$ with $s=\mathtt  j\nu$. 
Thus, its entry $H_{b\ell}(\mathtt j\nu)$ quantifies both the magnitude amplification and phase shift effects from the input fluctuation at load bus $\ell\in\ccalL$ to the output frequency deviation in bus $b$. The total frequency response at any bus $b$ results from the linear superposition of those individual responses due to multiple load buses.

\subsection{Load Fluctuation Model and System Frequency Response}\label{sec2-2}
To characterize FOs driven by large load fluctuations such as AI workloads, we consider several waveform models for $\mathbf{u}$. 
In addition to the sinusoidal waveform commonly used in FO studies, square and trapezoidal waveforms are considered to represent abrupt and ramp-limited variations\footnote{There are standards for maximum allowed active power ramp rate varying from 10--60 MW/min and 0.5 MW/s for various operators, such as Fingrid, American Transmission Company (ATC), and Alberta Electric System Operator (AESO) \cite{esig2026large}. In other regions, although there is no standard with specified ramp rate limit, requirements to limit the rapid ramping are being discussed in various ISOs and organizations, e.g., North American Electric Reliability Corporation (NERC), Electric Reliability Council of Texas (ERCOT), and California Independent System Operator (CAISO) \cite{nerc2025char, esig2026large, caiso2026}.}, respectively, as observed in practical large loads such as datacenters \cite{jensen2026easyrider, chen2025electricity}.

To provide a unified harmonic representation, we model each load fluctuation as an odd-symmetric real-valued periodic waveform expanded in Fourier series as
\begin{equation}
    u_{\ell}(t)=\rho_{\ell}\sum_{m=1}^{\infty}c_{\ell,m}\,\mathrm{sin}\big( m\left(\omega_{\ell} t+\phi_{\ell}\right)\big)
    \label{genwaveform}
\end{equation}
where $\omega_{\ell}$ is the fundamental fluctuation frequency of each source, with $\rho_{\ell}$ and $\phi_{\ell}$ respectively denoting its amplitude and phase offset. The waveform type can be specified by the Fourier coefficients $c_{\ell,m}$ as
\begin{equation}
    \begin{cases}
        \text{sinusoidal:} & c_{\ell,1}^{\mathrm{sin}}=1,\,c_{\ell,m}^{\mathrm{sin}}=0,~\forall m>1, \\
        \text{square:}     & c_{\ell,2m-1}^{\mathrm{sq}}=\frac{4}{\pi(2m-1)},\,c_{\ell,2m}^{\mathrm{sq}}=0\quad\forall m, \\
        \text{trapezoidal:} & c_{\ell,m}^{\mathrm{tr}}=c_{\ell,m}^{\mathrm{sq}}\,\mathrm{sinc}\Large( \frac{m\omega_{\ell}\rho_{\ell}}{r_{\ell}}\Large)\quad\forall m, \\
    \end{cases}
    \label{genwaveform2}
\end{equation}
where $m=\left[1,2,\cdots\right]$ denotes the harmonic order and $r_\ell$ is the ramp rate limit. This common waveform model underlies both deterministic and probabilistic FO analyses, which differ in how the waveform parameters are treated.

For the deterministic FO analysis, $\rho_\ell$, $\omega_\ell$, and $\phi_\ell$ are treated as deterministic parameters. To account for the spatial load distribution, we introduce a binary vector $\boldsymbol{\chi}_\ccalL$, where each element $\chi_\ell\in\{0,1\}$ indicates whether a fluctuating load exists in bus $\ell$. Accordingly, the set of active fluctuating loads is defined as $\ccalS=\{\ell\in\ccalL: \chi_{\ell}=1\}$. Substituting the fluctuation model \eqref{genwaveform} into the linearized grid dynamics \eqref{ltisys1}--\eqref{ltisys2} yields the steady-state frequency deviation at bus $b$:
\begin{equation}
    y_{b}(t)=\sum_{\ell\in\ccalS,m}\rho_{\ell}c_{\ell,m}\,\mathrm{Im}\left\{H_{b\ell}\left({\mathtt j}m\omega_{\ell}\right)e^{
    {\mathtt j}\left(m\left(\omega_{\ell}t+\phi_{\ell}\right)\right)}\right\}. 
    \label{sysresponse}
\end{equation}
Thus, if the fundamental or harmonic components overlap with high-gain regions of $|H_{b\ell}({\mathtt j}\omega)|$, particularly likely for weakly damped systems, they can induce large frequency oscillations.

For the probabilistic FO analysis, $\rho_\ell$ remains a prescribed deterministic amplitude, while $\omega_\ell$ and $\phi_\ell$ are treated as random variables. Specifically, we assume the fundamental frequency follows a uniform distribution as $\omega_{\ell}\sim\mathbb{U}(\Omega_{\mathrm{base}})$ with the range $\Omega_{\mathrm{base}}=\left[\omega_{\min},\omega_{\min}+\Delta_{\omega}\right]$. Here, $\omega_{\min}$ and $\Delta_{\omega}$ are the given parameters defining the lower bound and bandwidth of the frequency uncertainty, respectively. The frequency $\omega_\ell$ is modeled as a time-invariant random parameter; thus, its value is uncertain in the probabilistic model but remains constant over time once specified. Accordingly, the overall frequency band of interest $\Omega$ becomes 
\begin{equation}
    \Omega=\bigcup_{m}\Big[m\omega_{\min},m(\omega_{\min}+\Delta_{\omega})\Big].
    \label{prob_range}
\end{equation}
In addition, we assume each phase offset follows $\phi_{\ell}\sim\mathbb{U}[0,2\pi]$, with $\phi_\ell$ independent of $\omega_\ell$, and that the random parameters are mutually independent across the sources.

Under these assumptions, one can model $u_\ell(t)$ as a wide-sense stationary (WSS) process. In particular, averaging over the uniformly distributed phase yields $\mathbb{E}[u_\ell(t)]=0$ and eliminates the cross terms between distinct harmonics. Accordingly, its autocorrelation depends only on the time lag $\tau$ and is given by
\begin{equation}
    \begin{aligned}
    R^u_{\ell}(\tau)&:=\mathbb{E}\big[u_{\ell}(t)u_{\ell}(t+\tau)\big] \\
    &=\frac{\rho_{\ell}^{2}}{2}\sum_{m=1}^{\infty}\mathbb{E}\left[c_{\ell,m}^{2}\cos(m\omega_{\ell}\tau)\right].
    \end{aligned}
    \label{autocorr}
\end{equation}
The corresponding power spectral density (PSD) at any frequency $\nu$ is defined as:
\begin{equation}
    S^u_{\ell}(\nu) := \int_{-\infty}^{\infty}R^u_{\ell}(\tau)e^{-j\nu\tau}d\tau~.
    \label{psd1}
\end{equation}
As a result, for the general waveform in \eqref{genwaveform}, its PSD becomes
\begin{equation}
    S^u_{\ell}(\nu) = \frac{\pi\rho_{\ell}^{2}}{2}\sum_{m=1}^{\infty}\mathbb{E}\Big[c_{\ell,m}^{2}\{\delta(\nu-m\omega_{\ell})+\delta(\nu+m\omega_{\ell})\}\Big]
    \label{psd2}
\end{equation}
where $\delta(\cdot)$ denotes the Dirac delta function.
For sinusoidal and square waveforms, their $\{c_{\ell,m}\}$ coefficients are independent of $\omega_l$, yielding
\begin{equation}
    S^u_{\ell}(\nu) = \frac{\pi\rho_{\ell}^{2}}{2}\sum_{m=1}^{\infty}c_{\ell,m}^{2}\mathbb{E}\Big[\delta(\nu-m\omega_{\ell})+\delta(\nu+m\omega_{\ell})\Big].
    \label{psd3}
\end{equation}
For a given value of $\omega_\ell$, the PSD consists of discrete harmonic lines located at $m\omega_{\ell}$. Averaging over $\Omega_{\mathrm{base}}$ transforms each harmonic line into a continuous spectral component. For the trapezoidal waveform, the spectral distribution within each band is additionally shaped by the frequency-dependent coefficient $c_{\ell,m}^{\mathrm{tr}}$.  


For the linear time-invariant system, the probabilistic system response follows directly from the input PSDs. Since the random parameters are mutually independent across sources and each input has zero mean, the cross-spectra between distinct inputs vanish. We can then form the diagonal input PSD matrix as $\mathbf{S}^{\mathbf{u}}(\nu)=\mathrm{diag}\left(\chi_{1}S^{u_{1}}(\nu),\cdots,\chi_{n_\ell}S^{u_{n_\ell}}(\nu) \right)$. 
The corresponding output $\mathbf{y}(t)$'s PSD becomes $\mathbf{S}^{\mathbf{y}}(\nu):=\mathbf{H}({\mathtt j}\nu)\mathbf{S}^{\mathbf{u}}(\nu)\mathbf{H}^{*}({\mathtt j}\nu)$ with the superscript $*$ for the Hermitian transpose, and thus the bus-wise output PSD is given by the diagonal entry of $\mathbf{S}^{\mathbf{y}}(\nu)$, as 
\begin{equation}
    S^y_{b}(\nu)=\sum_{\ell\in\ccalS}|H_{b\ell}(\mathtt j\nu)|^{2}S^{u}_{\ell}(\nu).
    \label{busPSD}
\end{equation}
This PSD expression is the spectral counterpart of the deterministic FO response in \eqref{sysresponse}, explicitly showing that each source spectrum is weighted by the squared grid transfer gain $|H_{b\ell}(\mathtt j\nu)|^2$. 

\section{Oscillation Severity Assessment}\label{sec3}

This section develops analytical metrics for assessing the severity of FOs driven by distributed load fluctuations. Following the analysis in Sec.~\ref{sec2}, we characterize the oscillatory impact from both deterministic and probabilistic perspectives. The former evaluates the \textit{worst-case} frequency deviation in steady-state under the condition of \textit{coherent phase alignment}, while the latter quantifies the band-limited response variance caused by random and spectrally dispersed fluctuations. Together, these metrics provide effective FO assessments through the lens of worst-case coherent forcing or statistically dispersed load fluctuations.

\subsection{Deterministic Severity Metrics}\label{sec3-1}

Using the outputs in \eqref{sysresponse}, we first define the \textit{deterministic severity metric} as the largest attainable frequency deviation in \textit{steady-state} across all buses, forcing frequencies, and source phase offsets, as given by
\begin{equation}
    \ccalM_{\mathrm{det}}:=\underset{\omega, b, \boldsymbol{\phi}_{\ccalL}}{\mathrm{max}} \,\underset{t \geq T_{0}}{\mathrm{sup}} |y_b(t)|
    \label{gen_det_metric}
\end{equation}
by choosing a sufficiently large $T_0$ after all transient dynamics decay. This metric can be used to specify the worst-case FO level under a \textit{coherent forcing} condition, in which all fluctuation sources operate at a common fundamental frequency $\omega$ with phase shifts perfectly aligned with the system response.

\begin{figure}[t]
    \centering
    \begin{tabular}{cc}
    \includegraphics[width=0.47\columnwidth]{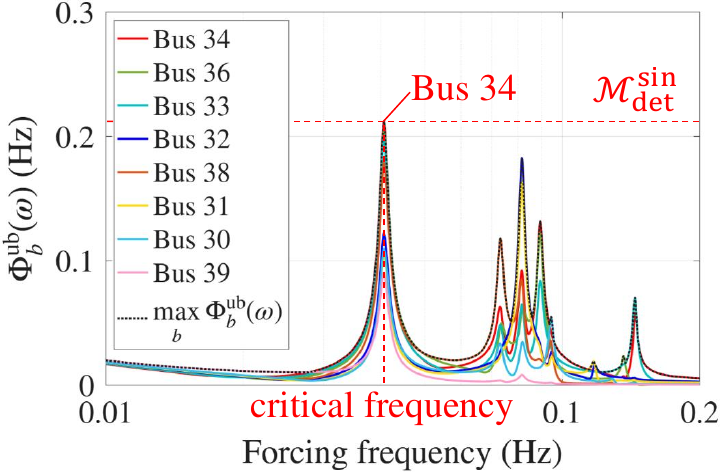} & \includegraphics[width=0.47\columnwidth]{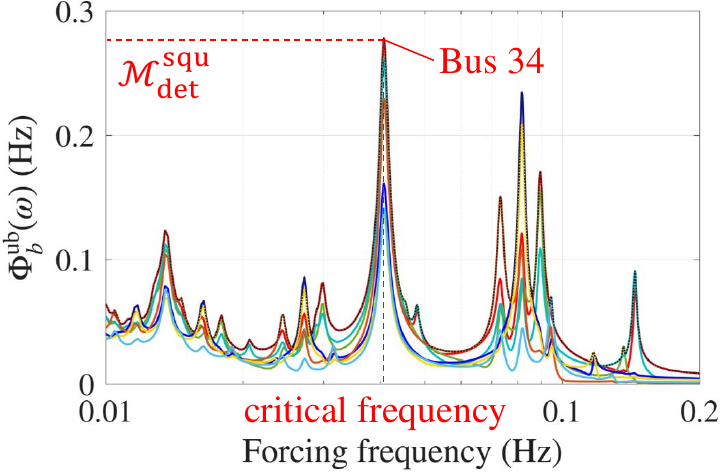}\\
    (a) & (b) \\
    \end{tabular}
    \caption{Deterministic amplification across output bus and forcing frequencies for (a) sinusoidal and (b) square waveforms in the IEEE 39-bus system.}
    \label{buswiseMdetsin}
    \vspace{-0.3cm}
\end{figure}

First, let us consider sinusoidal fluctuations for which the maximization in \eqref{gen_det_metric} over  $\boldsymbol{\phi}_{\ccalL}$ admits an exact analytical solution with given $\left(\omega,b\right)$.  By selecting $\phi_{\ell}^{*}=-\angle H_{b\ell}(\mathtt j\omega)+\phi_0$ for every load $\ell$ with an arbitrary common angle $\phi_0$, all sources can be aligned to the same angle in \eqref{sysresponse}. Under this \textit{phase alignment} setting, the contribution of any source $\ell$ to $y_b(t)$, i.e., each summand in \eqref{sysresponse}, 
shares the same angle, and can be directly aggregated over every $\ell$ without considering the phase angle effects. 
As for general non-sinusoidal waveforms, the exact maximization in \eqref{gen_det_metric} is intractable because the phase angles of harmonic components are related to each other and cannot be independently chosen to achieve phase alignment. To obtain a tractable gain, we omit this underlying phase coupling by applying the triangle inequality to the harmonic response in \eqref{sysresponse}, thereby bounding each harmonic contribution by its magnitude.
Accordingly, the deterministic source-to-output contribution at bus $b$ can be defined as
\begin{equation}
    F_{b\ell}^{\mathrm{det}}(\omega)=\rho_{\ell}\sum_{m}\left|c_{\ell,m}\right|\left|H_{b\ell}({\mathtt j}m\omega)\right|, \quad \forall\ell\in\ccalS.
    \label{fbl}
\end{equation}
Notice that \eqref{fbl} generalizes the sinusoidal case with $m=1$ only.
By aggregating the contributions from all forcing sources, the bus-wise upper-bound response is given by
\begin{equation}
    \Phi_{b}^{\mathrm{ub}}(\omega)=\sum_{\ell\in\ccalS}F_{b\ell}^{\mathrm{det}}(\omega).
    \label{ubresponse}
\end{equation}
The corresponding deterministic upper-bound (UB) severity metric is then defined as
\begin{equation}
    \ccalM_{\mathrm{det}}^{\mathrm{ub}}=\underset{\omega,b}{\mathrm{max}}\;\Phi_{b}^{\mathrm{ub}}(\omega)
    \label{sev_det}
\end{equation}
where its complete expression is provided in Table~\ref{seveqn}.
While mathematically meaningful, this bound could be overly conservative or practically unachievable for the non-sinusoidal waveforms, as it essentially requires the simultaneous time-alignment of all harmonic peaks.




\begin{table}[t]
\centering
\caption{Summary of Oscillation Severity Metrics} 
\label{seveqn}
\begin{tabular}{c|l}
\thickhline
Metric            & Mathematical formulation       \\ \thickhline
$\ccalM_{\mathrm{det}}^{\mathrm{ub}}$  
& $\underset{\omega,b}{\mathrm{max}}  \underset{\ell\in\ccalS}{\sum}  \left(\rho_{\ell}  \underset{m}{\sum}  \left|c_{\ell,m}\right|  \left|H_{b\ell}({\mathtt j}m\omega)\right| \right)$\\ \hline

$\ccalM_{\mathrm{det}}^{\mathrm{rms}}$
& $\underset{\omega,b}{\mathrm{max}}\sqrt{\frac{1}{2}\underset{m}{\sum} \Big(\underset{\ell\in\ccalS}{\sum}\rho_{\ell}\left|c_{\ell,m}\right|\left|H_{b\ell}({\mathtt j}m\omega)\right|\Big)^2}$ \\ \hline

$\ccalM_{\mathrm{prob}}$
& $\underset{b}{\mathrm{max}}\sqrt{\underset{\ell\in\ccalS}{\sum} \frac{\rho_{\ell}^2}{2\Delta_{\omega}} \Big(\underset{m}{\sum}\int_{\omega_{\mathrm{min}}}^{\omega_{\mathrm{min}}+\Delta_{\omega}}\left|c_{\ell,m}\right|^{2}\left|H_{b\ell}({\mathtt j}m\omega)\right|^{2}d\omega} \Big)$ \\ \thickhline
\end{tabular}
\vspace{-0.3cm}
\end{table}

To mitigate this conservativeness, we replace the direct summation of all harmonic peaks in \eqref{sysresponse} by the root-mean-square (RMS) aggregation. Specifically, we first aggregate the source contributions within each harmonic, then combine different harmonics in an RMS sense. The bus-wise RMS response profile can then be defined as
\begin{equation}
    \Phi_{b}^{\mathrm{rms}}(\omega)=\sqrt{\frac{1}{2}\sum_{m}\Big(\sum_{\ell\in\ccalS}\rho_{\ell}\left|c_{\ell,m}\right|\left|H_{b\ell}({\mathtt j}m\omega)\right|\Big)^2}.
\end{equation}
The corresponding RMS-based deterministic severity metric is
\begin{equation}
\ccalM_{\mathrm{det}}^{\mathrm{rms}}=\max_{\omega,b}\Phi_b^{\mathrm{rms}}(\omega),
\label{ns_rms}
\end{equation}
whose detailed expression is also summarized in Table~\ref{seveqn}. This RMS-based metric no longer needs the simultaneous peak alignment across the harmonics unlike \eqref{sev_det}; instead, it maintains a worst-case magnitude aggregation across sources within each harmonic while avoiding cross-harmonic peak alignment. It therefore serves as an upper bound on RMS severity and a less conservative surrogate for the deterministic FO severity under non-sinusoidal forcing.

Considering the IEEE 39-bus system as an example, Fig.~\ref{buswiseMdetsin} plots the bus-wise response profile in \eqref{ubresponse}, $\Phi_{b}^{\mathrm{ub}}(\omega)$, for selected buses. 
For this test system, the critical forcing frequency remains nearly identical
across buses and waveforms, due to the dominance of low-frequency electromechanical modes. In many other systems, however, high-order harmonics may excite more weakly damped modes. Related to this harmonic effect, Fig.~\ref{buswiseMdetsin}(b) shows that the square waveform yields the largest amplification at the critical frequency due to its high harmonic content. We have also tested the trapezoidal waveform with reduced amplification than the square one, followed by the sinusoidal waveform in Fig.~\ref{buswiseMdetsin}(a).


\subsection{Probabilistic Severity Metrics}\label{sec3-2}

Under random frequencies and phases for the fluctuations, the instantaneous peak response model in \eqref{gen_det_metric} is no longer suitable, and we resort to the spectral response in the output PSD $S^y_{b}(\nu)$ in \eqref{busPSD}. To account for the range of frequency  $\Omega$  defined in \eqref{prob_range}, we first define the probabilistic source-to-output contribution as
\begin{equation}
    F_{b\ell}^{\mathrm{prob}}=\frac{\chi_{\ell}}{\pi}\int_{\Omega} |H_{b\ell}({\mathtt j}\nu)|^{2}S^{u}_{\ell}(\nu)d\nu, \quad \forall\ell\in\ccalS.
    \label{fblprob}
\end{equation}
The band-limited response variance at bus $b$ can be obtained by aggregating the contributions from all forcing sources:
\begin{equation}
    J_{b} := \sum_{\ell}F_{b\ell}^{\mathrm{prob}}= \frac{1}{\pi}\int_{\Omega} S_{b}^{y}(\nu)d\nu, \label{probenergy}
\end{equation}
which aggregates all spectral components under the variation of $\omega_\ell$. Accordingly, its RMS counterpart, given by $\sqrt{J_{b}}$, shares the same physical unit as $y_b(t)$ and thus quantifies the average oscillation magnitude under the probabilistic setting, used to define the probabilistic severity metric as 
\begin{equation}
    \ccalM_{\mathrm{prob}}=\underset{b}{\mathrm{max}}\,\sqrt{J_{b}}~,\label{probmax}   \end{equation}
which is also detailed in Table~\ref{seveqn}.
Unlike deterministic metrics, $\ccalM_{\mathrm{prob}}$ no longer allows for the analysis of coherent phase alignment. Rather, as it quantifies the largest band-limited RMS magnitude under the given distribution of $\omega_\ell$, $\ccalM_{\mathrm{prob}}$ complements the deterministic ones by measuring the system vulnerability for uncertain $\omega_\ell$.

\begin{figure}[t]
    \centering
    \includegraphics[width=0.975\columnwidth]{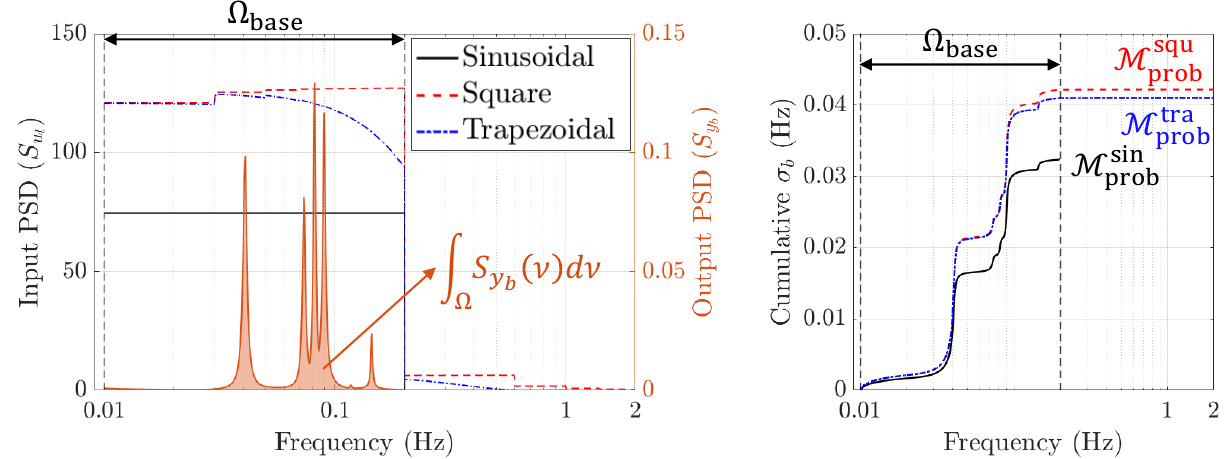}
    \caption{Example of input and output PSDs and the derivation of probabilistic severity metric in the IEEE 39-bus system.}
    \label{probiopsd}
    \vspace{-0.3cm}
\end{figure}

Fig.~\ref{probiopsd} illustrates the construction of $\ccalM_{\mathrm{prob}}$ from the input spectra and the grid frequency response in the IEEE 39-bus system. The input PSDs are weighted by the grid transfer gains in \eqref{busPSD}, and their contributions are accumulated into the band-limited response variance $J_b$ and hence $\ccalM_{\mathrm{prob}}$ through the maximization over output buses. For sinusoidal forcing, the input PSD is confined to $\Omega_{\mathrm{base}}$, while the non-sinusoidal waveforms introduce additional harmonic bands. When these harmonics overlap with high-gain frequency regions, they increase the shaded output-PSD area representing $J_b$ and can raise $\mathcal M_{\mathrm{prob}}$.


\section{Admissible Limits for Load Fluctuations} \label{sec4}

We further analyze the admissible magnitudes under a prescribed steady-state frequency threshold $\Delta f_{\mathrm{max}}^{\mathrm{ss}}$, based on the severity metrics developed in Sec.~\ref{sec3}. We consider three different formulations: a deterministic limit, its risk-aware extension, and a band-limited probabilistic limit. 
The risk-aware extension employs chance constraints (CC) and conditional value-at-risk (CVaR) to control violation probability and upper-tail severity under frequency uncertainty while retaining worst-case phase aggregation. 
Finally, the band-limited probabilistic limit incorporates both frequency and phase randomness through the PSD-based severity metric developed in Sec.~\ref{sec3-2}.

Throughout this section, all fluctuation sources are assumed to share a common magnitude $\rho$ for a fair allocation. Then, each formulation determines the maximum admissible magnitude $\rho_{\max}$ by imposing the corresponding severity or risk measure subject to $\Delta f_{\mathrm{max}}^{\mathrm{ss}}$. Of course, a source-specific non-uniform limit is also possible if certain types of loads can have more restrictive requirements.


\subsection{Deterministic Limit} \label{sec4-1}

Using the deterministic severity metrics $\ccalM_{\mathrm{det}}^{\kappa}$ in \eqref{sev_det} and \eqref{ns_rms}, where $\kappa\in\{\mathrm{ub}, \mathrm{rms}\}$ indexes the upper-bound and RMS-based formulations, we can directly determine the source magnitudes to satisfy the given threshold $\Delta f_{\mathrm{max}}^{\mathrm{ss}}$. 

Under these settings, any metric $\ccalM_{\mathrm{det}}^{\kappa}$ scales linearly with $\rho$. By defining the normalized level of $\Gamma_{\mathrm{det}}^{\kappa}:=\ccalM_{\mathrm{det}}^{\kappa} |_{\rho_{\mathrm{ref}}}$ with the reference fluctuation $\rho_{\mathrm{ref}}=1$, we have $\ccalM_{\mathrm{det}}^{\kappa}=\rho\Gamma_{\mathrm{det}}^{\kappa}$\footnote{For trapezoidal fluctuations, the Fourier coefficients depend on $\rho$ at a fixed physical ramp rate. In the normalized limit formulation, the reference waveform is defined at $\rho_{\mathrm{ref}}=1$, and its ramp characteristics and coefficients are held fixed for amplitude scaling.}. 
Accordingly, to achieve  $\ccalM_{\mathrm{det}}^{\kappa} \leq \Delta f_{\mathrm{max}}^{\mathrm{ss}}$, the deterministic allowable limit simply becomes
\begin{equation}
    \rho_{\mathrm{max}}^{\kappa}=\frac{\Delta f_{\mathrm{max}}^{\mathrm{ss}}}{\Gamma_{\mathrm{det}}^{\kappa}}.
    \label{lim_det}
\end{equation}
As discussed in Sec.~\ref{sec3-2}, the choice of $\kappa$ is critical and could affect the conservativeness of \eqref{lim_det}. Using the upper bound $\ccalM_{\mathrm{det}}^{\mathrm{ub}}$ is appropriate for a conservative screening, while the RMS one $\ccalM_{\mathrm{det}}^{\mathrm{rms}}$ provides a more practical steady-state criterion especially for non-sinusoidal sources.

\subsection{Risk-aware Limit Under Frequency Uncertainty} \label{sec4-2}

As \eqref{lim_det} is limited to a common fixed frequency, we extend the deterministic upper-bound metric in \eqref{sev_det} to account for source-wise uncertainty in $\omega_{\ell}$. Recall that each $\omega_\ell \sim\mathbb{U} \left (\Omega_{\mathrm{base}}\right)$, independent from other sources in the probabilistic model given in Sec.~\ref{sec2-2}. Let $\boldsymbol{\rho}_{\ccalS}$ collect the fluctuation magnitudes $\rho_\ell$ for $\ell\in\ccalS$. For a given $\boldsymbol{\omega}_{\ccalL}$, the frequency-dependent upper-bound response under source-specific fluctuation magnitudes is then given by using \eqref{fbl} as:
\begin{equation}
\begin{aligned}
R(\boldsymbol{\omega}_{\ccalL};\boldsymbol{\rho}_{\ccalS})
&:= \max_b \sum_{\ell\in\ccalS}F_{b\ell}^{\mathrm{det}}(\omega_\ell)\\
&= \max_b \sum_{\ell\in\ccalS}\rho_\ell \sum_m |c_{\ell,m}||H_{b\ell}({\mathtt j}m\omega_\ell)|.
\end{aligned}
\label{risk_response}
\end{equation}
Using the Fourier coefficients evaluated at the uniform reference magnitude $\rho_{\mathrm{ref}}=1$, we further define $R_{\mathrm{ref}}(\boldsymbol{\omega}_{\ccalL}):= R(\boldsymbol{\omega}_{\ccalL};
\rho_{\mathrm{ref}}\mathbf{1})$.
Note that $R_{\mathrm{ref}}(\boldsymbol{\omega}_{\ccalL})$ is independent of $\rho$, with its distribution induced by the prescribed uncertainty model of $\omega_\ell$. While the deterministic limit considers the most severe common forcing frequency, it does not account for the likelihood of different source-frequency combinations. We therefore extend the assessment to source-wise frequency uncertainty using risk-aware formulations, such as CC and CVaR approaches for random $\{\omega_\ell \}$. The CC provides a direct reliability criterion by limiting the probability of threshold violation \cite{charnes1959chance}, while CVaR further accounts for the severity of responses within the upper tail \cite{rockafellar2002conditional}. Accordingly, the two measures provide different levels of protection between a probability-based criterion and a fully worst-case treatment.

Specifically, the CC-based criterion is formulated as
\begin{equation}
    \mathbb{P}(\rho R_{\mathrm{ref}}(\boldsymbol{\omega}_{\ccalL})\leq\Delta f_{\mathrm{max}}^{\mathrm{ss}})\geq \alpha
    \label{cc}
\end{equation}
with a prescribed confidence level $\alpha$. Defining the $\alpha$-quantile of $R_{\mathrm{ref}}$ as $q_{\alpha}(R_{\mathrm{ref}})$, our CC-based limit becomes
\begin{equation}
    \rho_{\mathrm{max}}^{\mathrm{cc}}=\frac{\Delta f_{\mathrm{max}}^{\mathrm{ss}}}{q_{\alpha}(R_{\mathrm{ref}})}. \label{cc_lim}
\end{equation}
Since $q_{\alpha}(R_{\mathrm{ref}})$ could be difficult to obtain analytically, it is commonly estimated by empirical methods like Monte Carlo (MC) sampling. 

To better mitigate tail events, we also use the CVaR as the risk measure for $R(\boldsymbol{\omega}_{\ccalL})$, given by
\begin{equation}
    \mathrm{CVaR}_{\alpha}(R_{\mathrm{ref}}):=\underset{\eta}{\mathrm{min}}\left(\eta+\frac{1}{1-\alpha}\mathbb{E}[\mathrm{max}(R_{\mathrm{ref}}-\eta,0)]\right).
\end{equation}
Similar to CC, $\mathrm{CVaR}_{\alpha}(R_{\mathrm{ref}})$ is also estimated from MC sampling. As CVaR is known to satisfy positive homogeneity \cite{rockafellar2002conditional}, we can obtain another risk-aware limit as 
\begin{align}
    \mathrm{CVaR}_{\alpha}(\rho R_{\mathrm{ref}})&=\rho\mathrm{CVaR}_{\alpha}(R_{\mathrm{ref}})\leq\Delta f_{\mathrm{max}}^{\mathrm{ss}} \label{cvar} \\ 
    \rho_{\mathrm{max}}^{\mathrm{cvar}}&=\frac{\Delta f_{\mathrm{max}}^{\mathrm{ss}}}{\mathrm{CVaR}_{\alpha}(R_{\mathrm{ref}})}. \label{cvar_limit}
\end{align}
This limit is generally smaller than the CC-based one in \eqref{cc_lim}, because the CVaR measure is known to be more conservative in penalizing the worst-case performance. 

\subsection{Band-based Probabilistic Limit} \label{sec4-3}

Last, we consider the probabilistic limit using the PSD-based metric in Sec.~\ref{sec3-2}. In addition to the frequency uncertainty, this limit further incorporates the random phase shifts and is thus less conservative than Sec.~\ref{sec4-2}. 

Recall that the input PSDs in \eqref{psd2} and \eqref{psd3} scale quadratically with $\{\rho_\ell\}$. Accordingly, for a uniform source magnitude $\rho$, the output PSD satisfies $S^y_{b}(\nu)=\rho^{2}\Sigma_{y_b}^{(1)}(\nu)$ with the normalized level $\Sigma_{y_b}(\nu)$ under unity $\rho$. Accordingly, the probabilistic severity metric in \eqref{probmax} also scales linearly with $\rho$, as 
\begin{equation}
    \ccalM_{\mathrm{prob}}=\rho\Gamma_{\mathrm{prob}} ~\mathrm{with}~ \Gamma_{\mathrm{prob}}:=\underset{b}{\mathrm{max}}\sqrt{\frac{1}{\pi}\int_{\Omega}\Sigma^y_{b}(\nu)d\nu}.
\end{equation}

Since $\ccalM_{\mathrm{prob}}$ represents an RMS-based measure, while the frequency threshold $\Delta f_{\mathrm{max}}^{\mathrm{ss}}$ specifies the peak deviation, we use the peak-to-RMS conversion for $\Delta f_{\mathrm{max}}^{\mathrm{ss}}$ based on the crest factor denoted by $\mathrm{CF}$ \cite{boyd1986multitone}. One common approach to determine $\mathrm{CF}$ is based on the input-side analysis, but this could be insufficient to represent the exact output-side ratio due to the system responses. To avoid repeated waveform-specific calibrations, we estimate a common output $\mathrm{CF}$ from unit-amplitude sinusoidal responses, reducing computational burden while providing a consistent peak-to-RMS conversion across the considered waveforms. 
For each MC sample $r$, we define the worst steady-state peak response across buses by
\begin{equation}
    Z_{r}^{\mathrm{sin}}=\underset{b}{\mathrm{max}}\;\underset{t\geq T_{0}}{\mathrm{sup}} \left|\Upsilon_{r}^{b,\mathrm{sin}}(t) \right|, ~\forall r
\end{equation}
with $\Upsilon_{r}^{b,\mathrm{sin}}(t)$ denoting the corresponding frequency deviation trajectory. 
This defines the quantile-based output $\mathrm{CF}$ as
\begin{equation}\mathrm{CF}_{\alpha}^{\mathrm{sin}}:=\frac{q_{\alpha}\left(Z_{r}^{\mathrm{sin}} \right)}{\Gamma_{\mathrm{prob}}^{\mathrm{sin}}},
\end{equation}
which provides an empirical mapping between the RMS-based metric and the corresponding peak response. This way, the probabilistic limit becomes
\begin{equation}
    \rho_{\mathrm{max}}^{\mathrm{prob}}=\frac{\Delta f_{\mathrm{max}}^{\mathrm{ss}}}{\mathrm{CF}_{\alpha}^{\mathrm{sin}}\Gamma_{\mathrm{prob}}}.
\end{equation}
This limit uses a statistically calibrated criterion to accommodate random phase shifts, providing a probabilistic estimate of admissible fluctuation magnitudes..

\section{Security-Constrained (SC) Optimization of \\ Load Fluctuations}\label{sec5}

The admissible limits in Sec.~\ref{sec4} assume a uniform fluctuation magnitude across all sources. However, in practical operation, individual large loads may exhibit different fluctuation magnitudes depending on their operating conditions. Therefore, we determine the source-wise optimization of fluctuation magnitudes to satisfy a prescribed frequency-security requirement under fixed load locations.

Let $\Psi(\boldsymbol{\rho}_{\ccalS})$ stand for the corresponding security measure, selected from the criteria developed in Sec.~\ref{sec4}. Under a given frequency-security requirement and individual source limits, one can maximize the aggregate fluctuation magnitudes, as
\begin{subequations}  \label{eq:opt}
    \begin{equation} \label{optrho}
        \boldsymbol{\rho}_{\ccalS}^{*} = ~\underset{\boldsymbol{\rho}_{\ccalS}}{\mathrm{argmax}} \sum_{\ell\in\ccalS}{\rho_{\ell}} 
    \end{equation}
    \begin{equation} \label{optcon1}
        \mathrm{s.t.} 
        ~\Psi(\boldsymbol{\rho}_{\ccalS})\leq \Delta f_{\max}^{\mathrm{ss}},
    \end{equation}
    \begin{equation} \label{optcon2}
        \underline{\rho}_{\ell} \leq \rho_{\ell} \leq \bar{\rho}_{\ell}, \; \forall\ell\in\ccalS
    \end{equation}
\end{subequations}
where 
\eqref{optcon1} enforces the prescribed frequency-security requirement, and \eqref{optcon2} imposes the known lower and upper fluctuation limits for each source.

The complexity of solving \eqref{eq:opt} depends on the adopted security metric and fluctuation waveform, as summarized in Table~\ref{opt}. For sinusoidal and square waveforms, $c_{\ell,m}$ is independent of $\rho_\ell$, such that the metrics in Table \ref{seveqn} consist of either linear functions or their Euclidean norms of $\boldsymbol{\rho}_{\ccalS}$. Accordingly, using the deterministic $\ccalM_{\mathrm{det}}^{\mathrm{ub}}$ and $\ccalM_{\mathrm{det}}^{\mathrm{rms}}$ yields linear and quadratic constraints, respectively resulting in a linear program (LP) and a quadratically constrained quadratic program (QCQP). For the risk-aware counterparts of $\ccalM_{\mathrm{det}}^{\mathrm{ub}}$ as discussed in Sec.~\ref{sec4-2}, we consider the general response
$R(\boldsymbol{\omega}_{\ccalL};\boldsymbol{\rho}_{\ccalS})$ in \eqref{risk_response} under source-specific fluctuation magnitudes. Using MC samples in Sec.~\ref{sec4-2}, the CC formulation is approximated as a mixed-integer LP (MILP) by introducing binary variables to indicate scenario violations. Meanwhile, the empirical CVaR constraint admits an equivalent linear reformulation, resulting in an LP. For the probabilistic metric in \eqref{probmax}, it is linearly related to the squared fluctuation magnitudes, namely $\epsilon_\ell=\rho_\ell^2$. Thus, we instead replace the objective in \eqref{optrho} with $\sum_{\ell\in\ccalS}\rho_\ell^2$, leading to an equivalent LP formulation, as both the new objective and $J_b$ are linear in $\epsilon_\ell$. 

\begin{table}[t]
\begin{threeparttable}
\centering
\caption{SC Optimization Formulation/Method}
\label{opt}
\begin{tabular}{l|l}
\thickhline
\begin{tabular}[c]{@{}l@{}} Security criterion\\ or waveform\end{tabular} & Optimization formulation or method \\ \thickhline

$\ccalM_{\mathrm{det}}^{\mathrm{ub}} \leq \Delta f_{\max}^{\mathrm{ss}} $  & Linear program (LP)\\ \hline
$\ccalM_{\mathrm{det}}^{\mathrm{rms}} \leq \Delta f_{\max}^{\mathrm{ss}}$ & Quadratically constrained QP (QCQP)  \\ \hline
$\mathbb{P}\left(R \leq \Delta f_{\max}^{\mathrm{ss}}\right)\geq\alpha$ & Mixed-integer LP (MILP)  \\ \hline
$\mathrm{CVaR}_{\alpha}\left(R\right) \leq \Delta f_{\max}^{\mathrm{ss}}$ & LP \\ \hline
$\ccalM_{\mathrm{prob}}\leq\Delta f_{\mathrm{max}}^{\mathrm{ss}}$ & LP in the variables $\epsilon_{\ell}=\rho_{\ell}^2$   \\ \hline \hline
Trapezoidal waveform  & Multi-start SQP with nonlinear constraints  \\ \thickhline
\end{tabular}
\begin{tablenotes}\footnotesize
      \item For compactness, $R$ denotes $R(\boldsymbol{\omega}_{\ccalL};\boldsymbol{\rho}_{\ccalS})$ defined in \eqref{risk_response}.
    \end{tablenotes}
\end{threeparttable}
\vspace{-0.3cm}
\end{table}

As for trapezoidal waveforms, they present significant issues in solving \eqref{eq:opt}, as their Fourier coefficients $\{c_{\ell,m}^{\mathrm{tr}}\}$ depend on $\rho_\ell$. This dependence makes the security constraint highly nonlinear, regardless of the aforementioned metrics. Therefore, the resulting optimization problems are solved using a generic nonlinear solver, the sequential quadratic programming (SQP) \cite{bonnans2006numerical} with a multi-start strategy \cite{marti2013multi}, by screening multiple feasible points for the initialization to  mitigate local solutions.

\section{Case Studies} \label{sec6}

\begin{figure}[t]
    \centering
    \begin{tabular}{cc}
    \includegraphics[width=0.44\columnwidth]{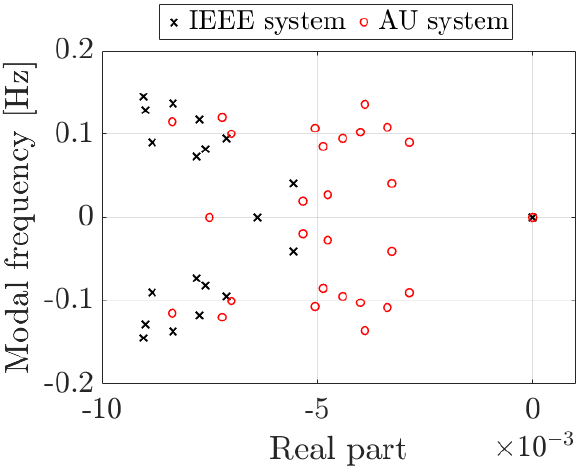} & \includegraphics[width=0.44\columnwidth]{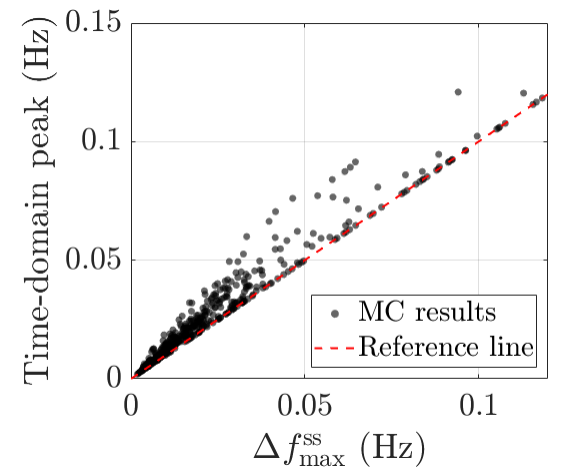}\\
    (a) & (b) \\
    \end{tabular}
    \caption{Results of (a) the eigen-analysis of two test systems and (b) the comparison of time-domain and steady-state peaks.}
    \label{eig}
    \vspace{-0.3cm}
\end{figure}

We evaluate the proposed methodology using two test cases: i) IEEE 39-bus system; and ii) 14-generator Australian power system. We first consider the eigen-analysis for the simplified models of both systems shown in Fig.~\ref{eig}(a). In the IEEE 39-bus system, the dominant low-frequency response is concentrated around a relatively narrow modal region, while the Australian system contains multiple weakly damped modes distributed over a broader range below approximately 0.2~Hz. The former provides a relatively simple benchmark dominated by a limited set of modes, while the latter exhibits more complex multimodal dynamics. Accordingly, we have set the frequency range $\Omega_{\mathrm{base}}=[0.01,0.2]~\mathrm{Hz}$ with the lower bound also consistent with real-world data \cite{choukse2025power,ko2026wide}. The two systems have approximately 6-GW and 22-GW capacity, respectively, and per the sizing difference, we have selected 4 fluctuating sources in the former and 7 sources in the latter. In addition, we have initially selected electrically distributed load buses across the system. For each source, its baseline magnitude is assumed to be 300 MW, with a 400 MW/sec ramp rate for the trapezoidal waveform. 
For the risk-constrained metrics and MC simulations, 2000 simulation runs have been adopted.

Our studies further set $\Delta f_{\max}^{\mathrm{ss}}=0.02$~Hz\footnote{Although no generally accepted standard or limit has yet been established for large load fluctuations, normal operating frequency bands adopted by major system operators are generally on the order of $\pm 0.05$--$\pm 0.2$~Hz around the nominal frequency \cite{entsoe, aemc}. Accordingly, the selected 0.02 Hz threshold provides a deliberately conservative criterion for sustained oscillatory deviations.} as the steady-state frequency-deviation threshold. Fig.~\ref{eig}(b) relates this threshold to the full time-domain peak deviations. The steady-state peak is evaluated after the transient response has decayed, i.e., over $t\geq T_0$, while the full time-domain peak is the maximum absolute deviation over the entire simulation horizon.
On average, the time-domain peak is approximately 1.42 times the steady-state one, implying that the resultant limit for time-domain peaks is around 0.03 Hz. 

\begin{figure}[t]
    \centering
    \begin{tabular}{cc}
    \includegraphics[width=0.45\columnwidth]{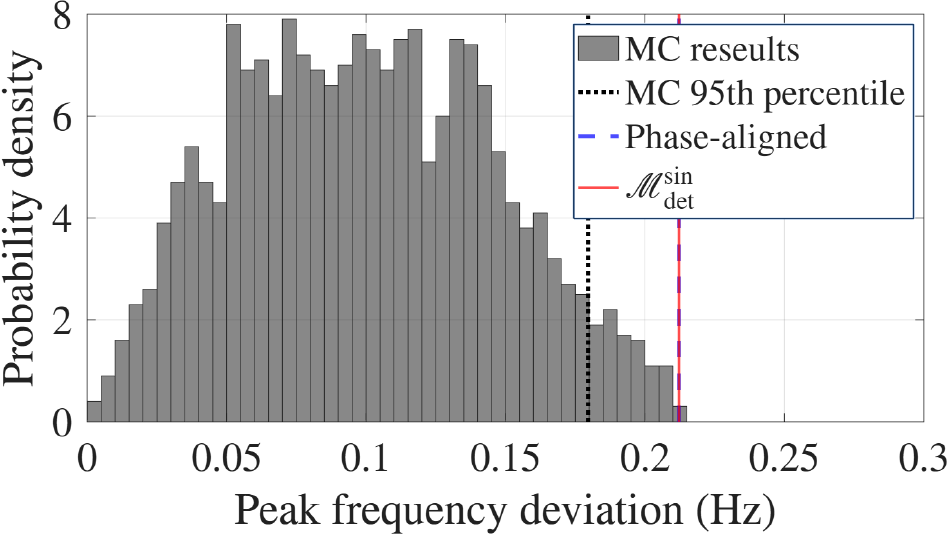} & \includegraphics[width=0.45\columnwidth]{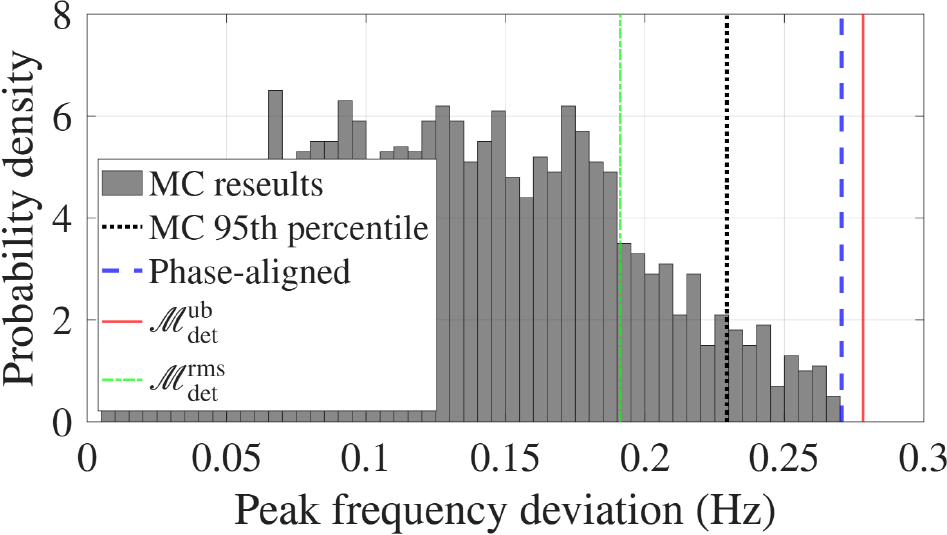}         \\
    (a)      &       (b) \\
    \includegraphics[width=0.45\columnwidth]{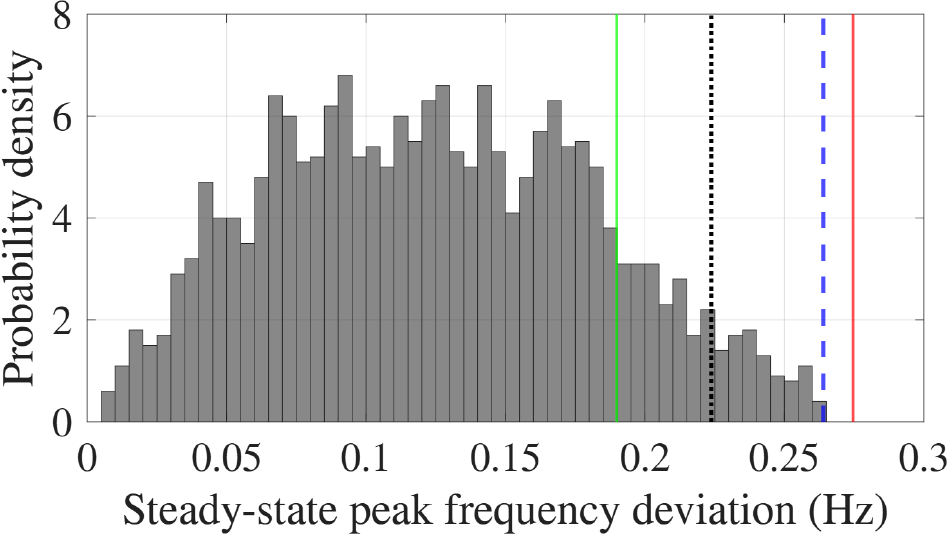} & \includegraphics[width=0.45\columnwidth]{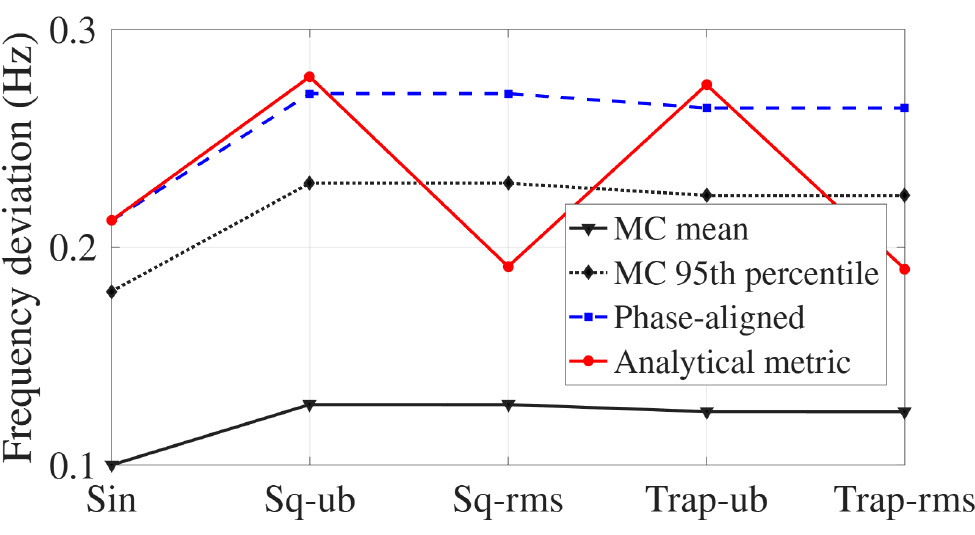}         \\
    (c)      &       (d)
    \end{tabular}
    \caption{Comparison of deterministic severity metrics and MC results in IEEE 39-bus system for (a) sinusoidal, (b) square, (c) trapezoidal waveforms, and (d) all waveforms.}
    \label{mcsev}
    \vspace{-0.3cm}
\end{figure}

\subsection{Validation of Oscillation Severity Metrics}

We first validate the deterministic metrics $\ccalM_{\mathrm{det}}$, using the IEEE 39-bus system. As shown in Fig.~\ref{buswiseMdetsin}, Bus~34 exhibits the largest response, with the critical forcing frequencies near 0.04~Hz across all three waveforms, consistent with the modal characteristics in Fig.~\ref{eig}(a). Fig.~\ref{mcsev} compares $\ccalM_{\mathrm{det}}$ with the corresponding time-domain simulations. The forcing frequency is fixed at the critical frequency identified for each waveform, while the phase offsets are randomly sampled. The square waveform produces the largest response, followed by trapezoidal and sinusoidal waveforms. For the sinusoidal case, the analytical metric closely matches the phase-aligned time-domain response, confirming the exact phase maximization. For the non-sinusoidal cases, $\ccalM_{\mathrm{det}}^{\mathrm{ub}}$ provides a conservative peak bound because harmonic contributions are independently maximized, while $\ccalM_{\mathrm{det}}^{\mathrm{rms}}$ relaxes this cross-harmonic peak alignment and falls between the MC mean and 95th percentile.

\begin{figure}[t]
    \centering
    \begin{tabular}{cc}
    \includegraphics[width=0.46\columnwidth]{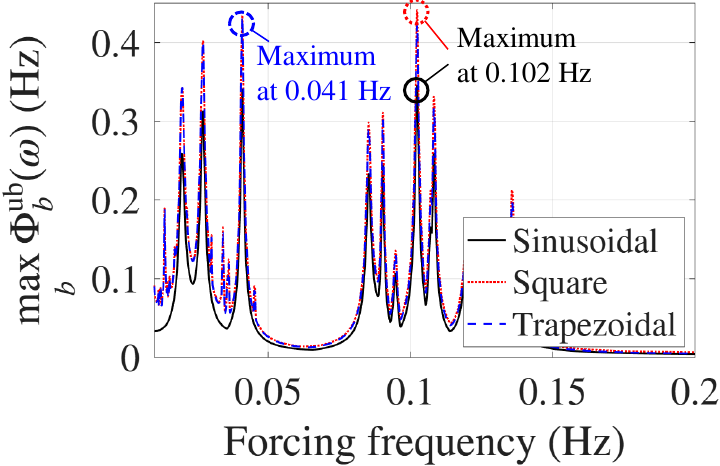} & \includegraphics[width=0.46\columnwidth]{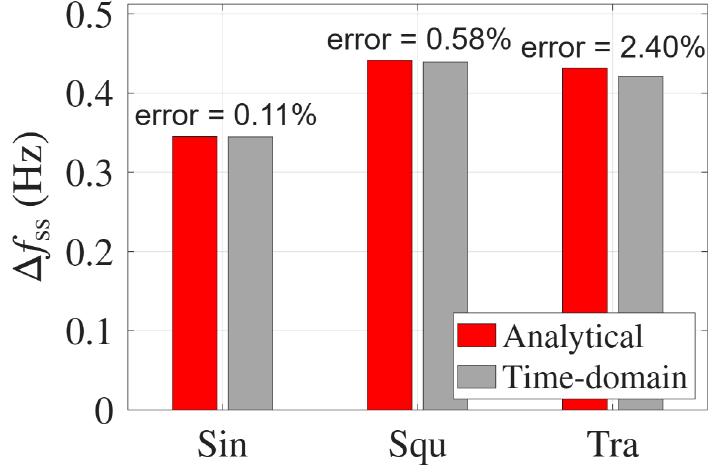}\\
    (a) & (b) \\
    \end{tabular}
    \caption{Deterministic severity assessment in the 14-generator Australian system: (a) frequency-dependent response profiles and (b) analytical versus time-domain responses.}
    \label{ausevdet}
    \vspace{-0.3cm}
\end{figure}

The Australian system provides a more challenging case since the critical forcing frequency and output bus vary with waveform characteristics. Fig.~\ref{ausevdet}(a) shows $\max_b\Phi_b^{\mathrm{ub}}(\omega)$ over the forcing-frequency range. The sinusoidal and square-wave responses peak near 0.102~Hz at Bus 101, while the trapezoidal response peaks near 0.041~Hz at Bus 302. 
This difference results from the frequency-dependent harmonic weighting of the trapezoidal waveform, which changes the relative excitation of the system modes.
Despite these differences, Fig.~\ref{ausevdet}(b) shows close agreement between the analytical and time-domain responses evaluated at the respective critical frequencies. These results demonstrate that waveform characteristics can alter not only oscillation severity but also the governing frequency and output location, emphasizing the importance of waveform-aware frequency-response assessment.

\begin{figure}[t]
    \centering
    \begin{tabular}{cc}
    \includegraphics[width=0.46\columnwidth]{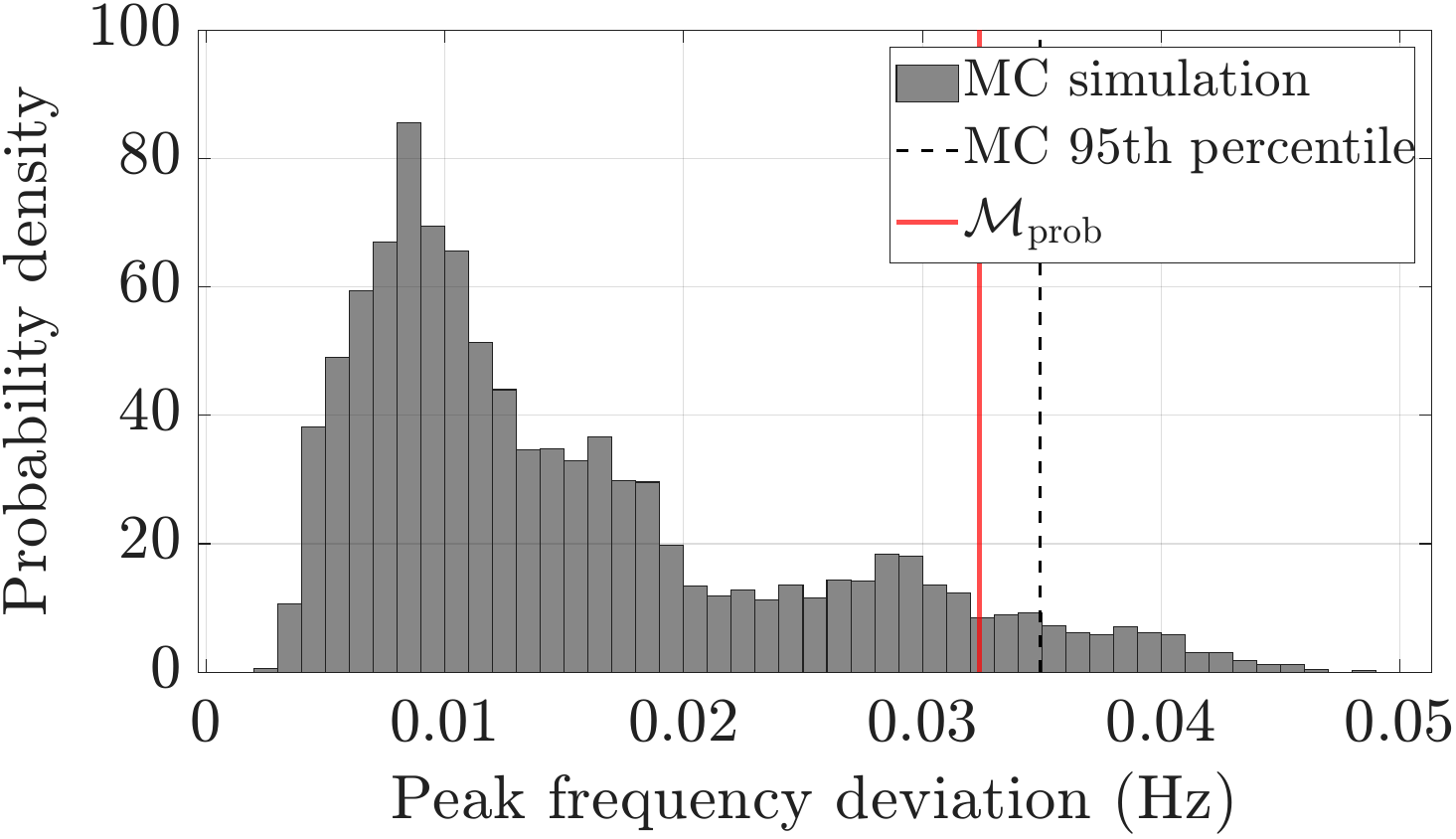} & \includegraphics[width=0.46\columnwidth]{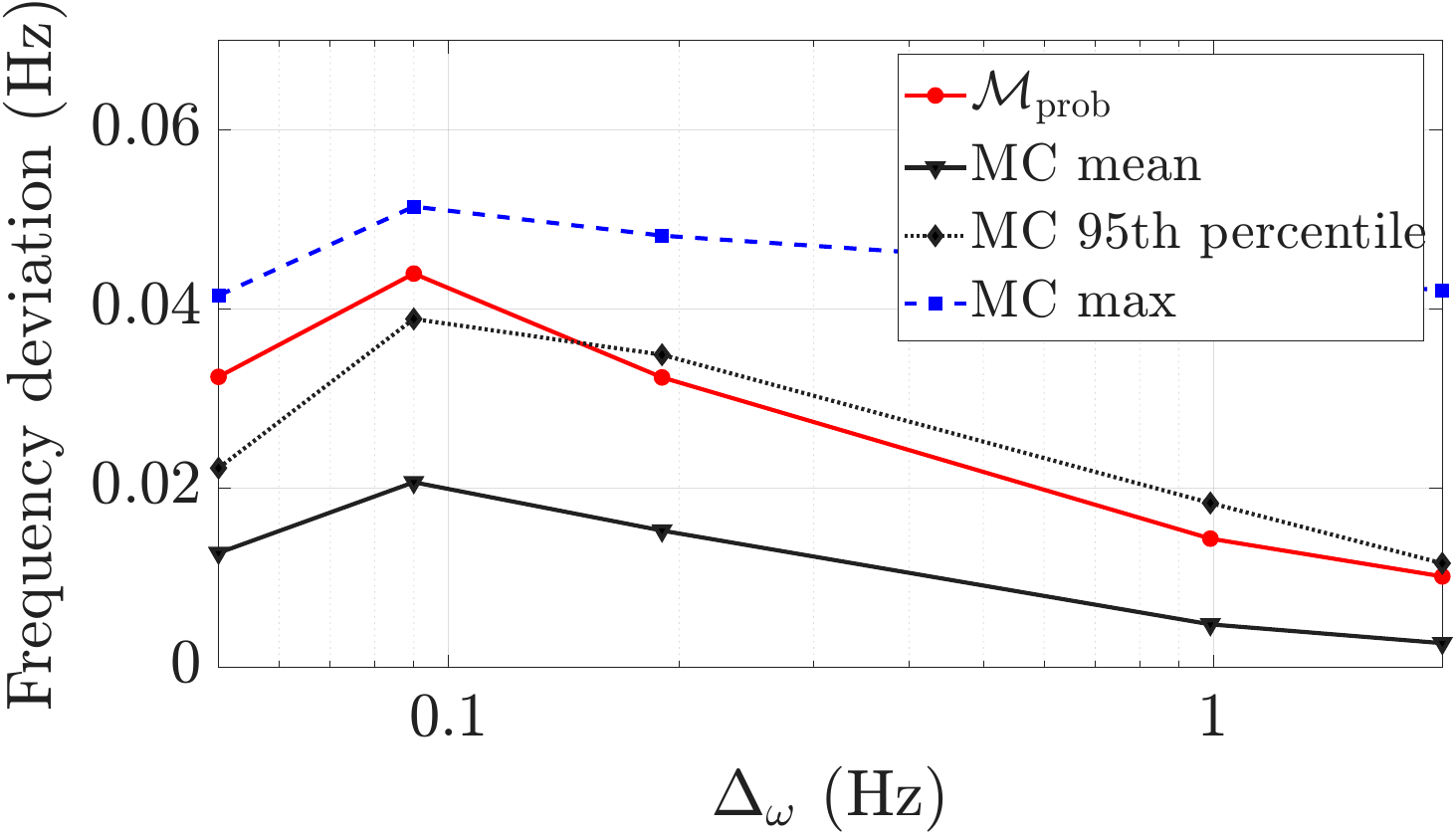}         \\
    (a)      &       (b) \\
    \includegraphics[width=0.46\columnwidth]{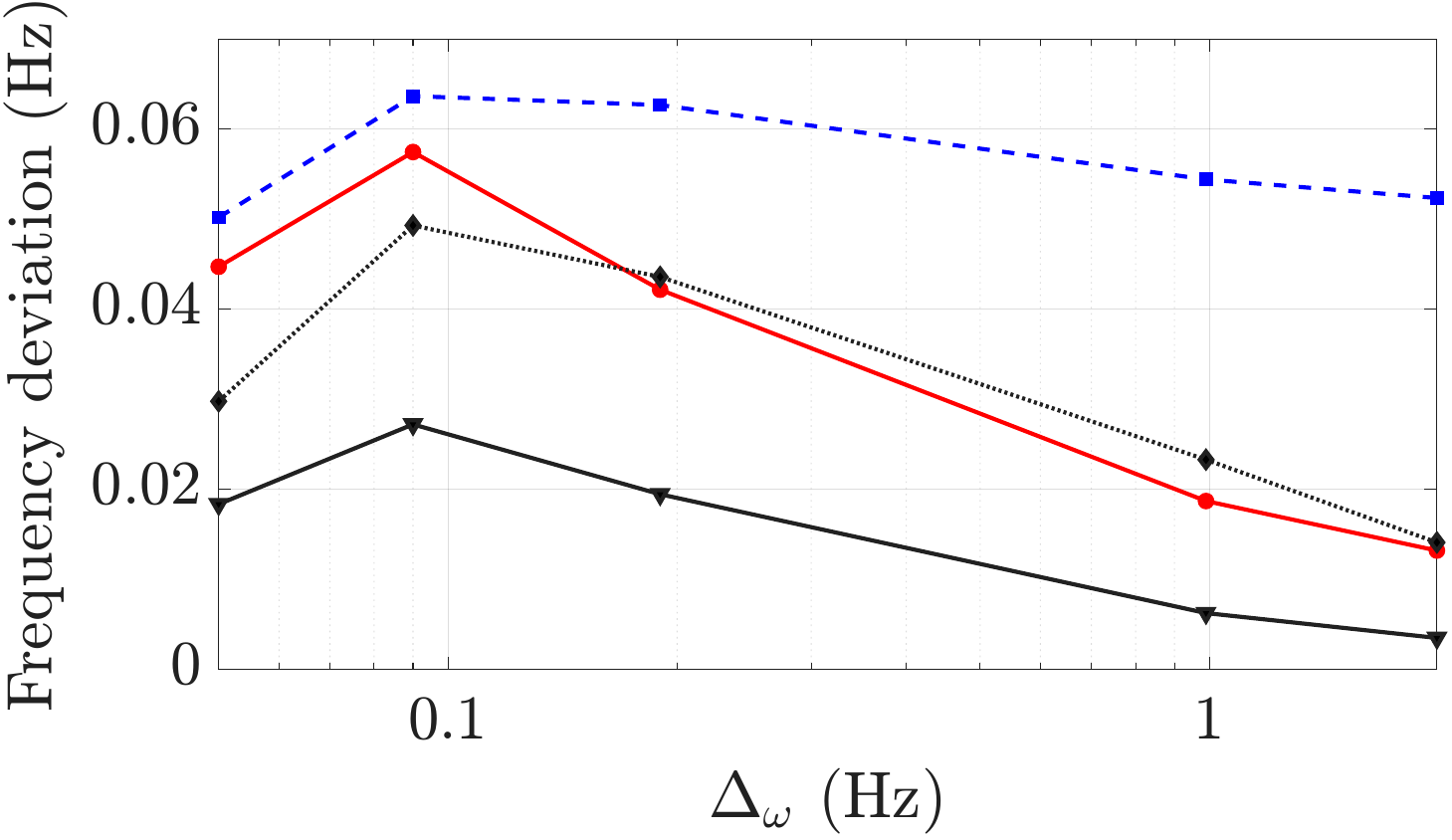} & \includegraphics[width=0.46\columnwidth]{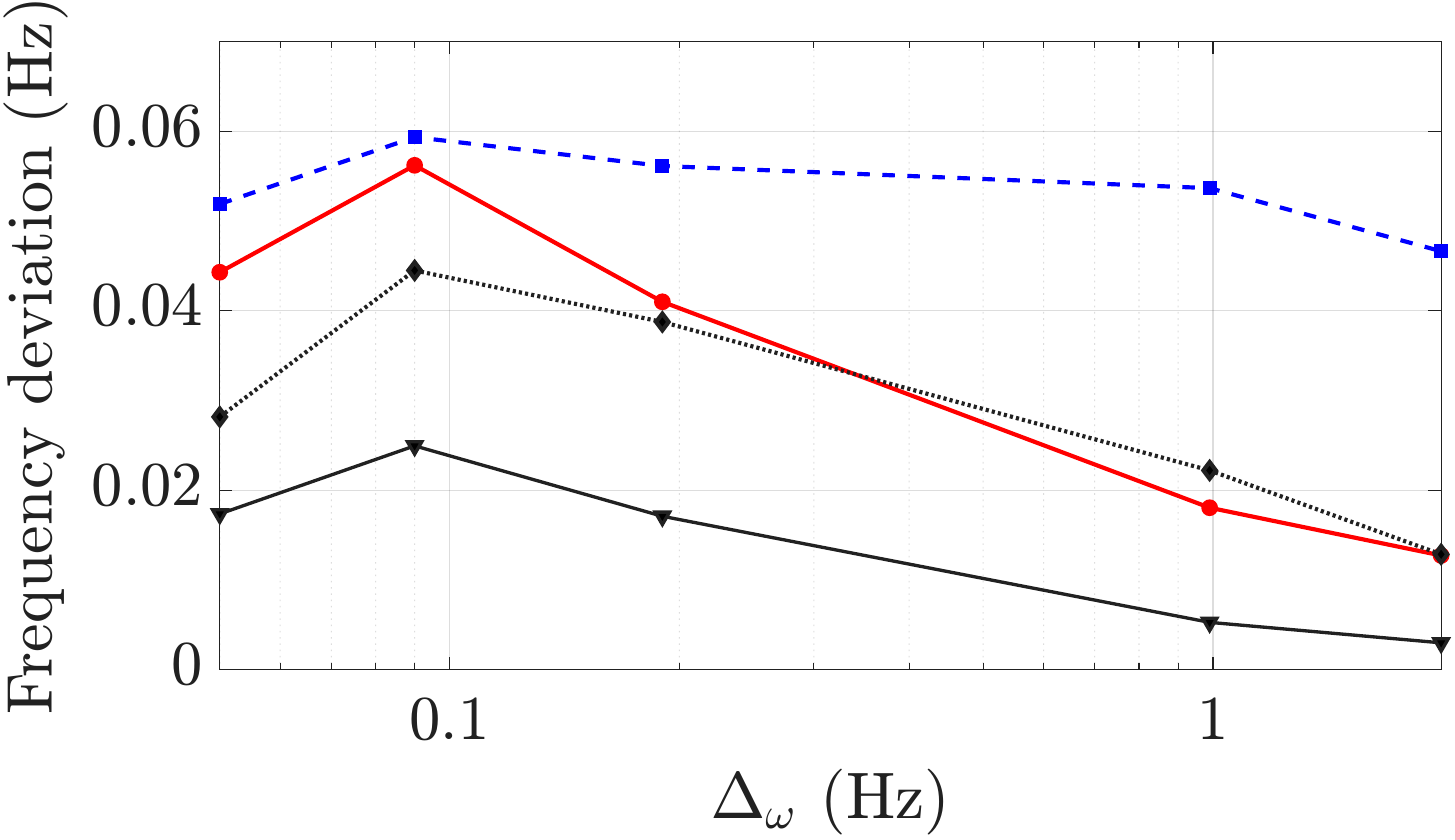}         \\
    (c)      &       (d)
    \end{tabular}
    \caption{Probabilistic severity assessment in the IEEE 39-bus system: (a) Comparison of metric and MC results for sinusoidal waveform and the impact of base frequency range width for (b) sinusoidal, (c) square, and (d) trapezoidal waveforms.}
    \label{probsev}
    \vspace{-0.3cm}
\end{figure}

\begin{figure}[t]
    \centering
    \includegraphics[width=0.7\columnwidth]{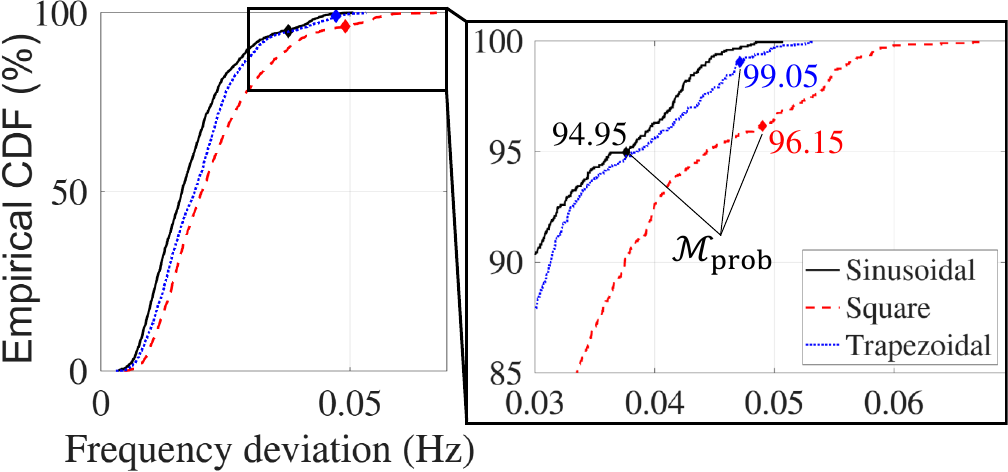}
    \caption{Comparison of the empirical CDF and probabilistic severity metrics in the Australian system.}
    \label{ausevprob}
    \vspace{-0.3cm}
\end{figure}

We next validate $\ccalM_{\mathrm{prob}}$ under independently sampled forcing frequencies $\omega_{\ell} \in \Omega_{\mathrm{base}}$ and phases $\phi_{\ell} \in [0,2\pi]$. Figs.~\ref{probsev} and \ref{ausevprob} show that $\ccalM_{\mathrm{prob}}$ consistently lies near or above the 95th percentile of the MC responses.
Moreover, Fig.~\ref{probsev}(a) shows that the MC-based frequency distribution for sinusoidal waveforms concentrates on the small values, as a result of random forcing frequencies and phase angles. 
Thus, $\ccalM_{\mathrm{prob}}$ provides a conservative spectral indicator of stochastic oscillation severity without requiring an exhaustive search over frequency and phase combinations. 
Furthermore, as shown in Fig.~\ref{probsev}(b), increasing $\Delta_{\omega}$ with a fixed $\omega_{\min}=0.01$ Hz generally reduces the $\ccalM_{\mathrm{prob}}$,  since a wider $\Delta_{\omega}$ beyond the weakly-damped modes likely gets less amplified by the grid. This is also consistent with Fig.~\ref{probiopsd}, where the dominant contribution is concentrated below approximately 0.2 Hz.

\begin{figure}[t]
    \centering
    \begin{tabular}{cc}
    \includegraphics[width=0.46\columnwidth]{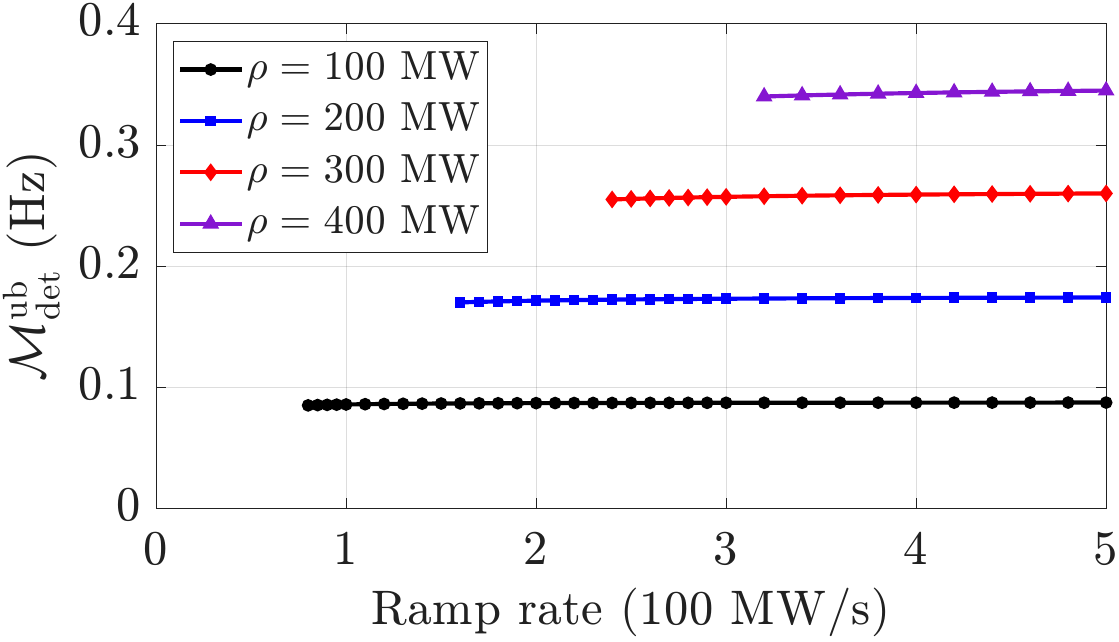} & \includegraphics[width=0.46\columnwidth]{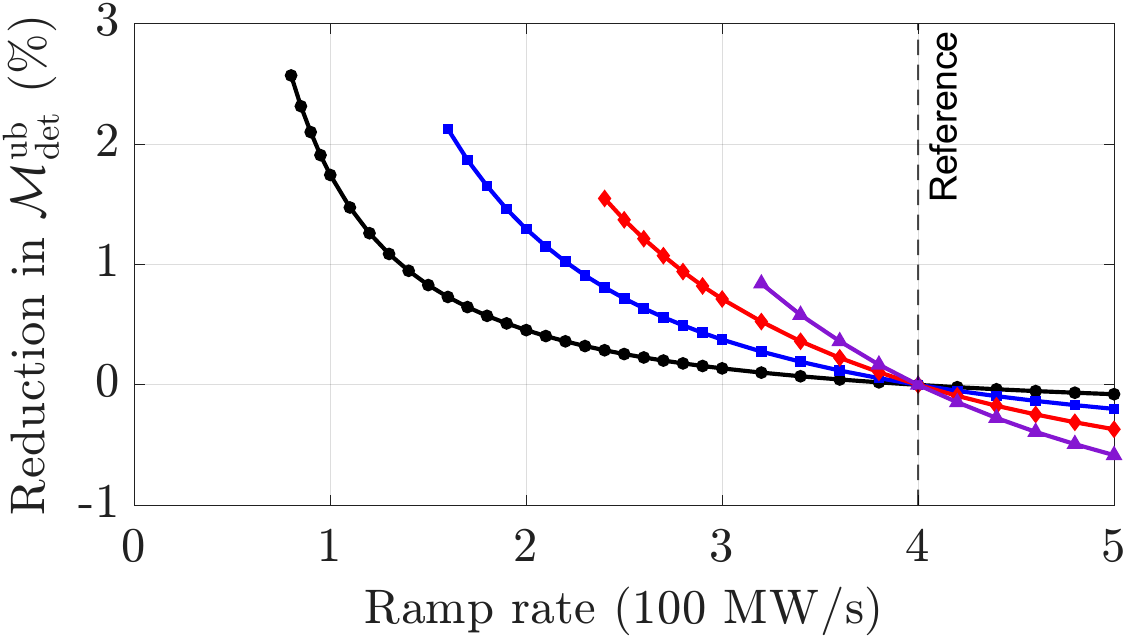}\\
    (a) & (b) \\
    \end{tabular}
    \caption{Impact of trapezoidal ramp rates on oscillation severity in the IEEE 39-bus system: (a) deterministic UB severity and (b) relative severity reduction.}
    \label{trapramp}
    \vspace{-0.3cm}
\end{figure}

Last, we analyze the effect of trapezoidal ramp rates on $\ccalM_{\mathrm{det}}^{\mathrm{ub}}$ in the IEEE 39-bus system, as illustrated in Fig.~\ref{trapramp}. We consider trapezoidal fluctuations ranging from 100 to 400 MW and vary the ramp rate within its feasible range. As expected, the severity increases with $\rho$ and generally increases with the ramp rate, with more pronounced changes near the minimum feasible ramp rate. The smaller relative reductions at larger $\rho$ reflect their higher minimum feasible ramp rates, which narrow the available range for ramp rate reduction. Meanwhile, the severity approaches a nearly constant level as the ramp rate increases, indicating a diminishing sensitivity to further ramp rate changes. These results highlight the importance of jointly considering fluctuation magnitudes and ramp rates when establishing grid-code requirements for oscillation mitigation.

\subsection{Validation of Admissible Limits}

We next evaluate the admissible fluctuation limits derived in Sec.~\ref{sec4} and validate the corresponding boundaries using time-domain MC simulations. All criteria use the same steady-state frequency threshold, $\Delta f_{\max}^{\mathrm{ss}} = 0.02$~Hz, with $\alpha=0.95$ for the CC and CVaR formulations. The MC validation follows the uncertainty assumption of each criterion. For the deterministic limits, we fix the forcing frequency at the waveform-specific critical frequency while randomly sampling the source phases. For the CC, CVaR, and probabilistic limits, both the source frequencies and phases are independently sampled according to Sec.~\ref{sec2-2}. Table~\ref{admlimit} summarizes the resulting admissible limits and time-domain violation rates for both test systems.

\begin{table}[t]
\centering
\begin{threeparttable}
\caption{Admissible Fluctuation Limit and Time-domain Violation Rate}
    \begin{tabular}{c|cc|cc|cc}
    \thickhline
    \multirow{2}{*}{}
    & \multicolumn{2}{c|}{Sinusoidal}
    & \multicolumn{2}{c|}{Square}
    & \multicolumn{2}{c}{Trapezoidal} \\ \cline{2-7}
    & \begin{tabular}[c]{@{}c@{}} Limit \\ (MW) \end{tabular}
    & \begin{tabular}[c]{@{}c@{}} Viol. \\ (\%) \end{tabular}
    & \begin{tabular}[c]{@{}c@{}} Limit \\ (MW) \end{tabular}
    & \begin{tabular}[c]{@{}c@{}} Viol. \\ (\%) \end{tabular}
    & \begin{tabular}[c]{@{}c@{}} Limit \\ (MW) \end{tabular}
    & \begin{tabular}[c]{@{}c@{}} Viol. \\ (\%) \end{tabular} \\ \hline
    \multicolumn{7}{c}{IEEE 39-bus system} \\ \hline
    Det-UB  & 28.3 & 0.00 & 21.6 & 0.00 & 21.6 & 0.00 \\
    Det-RMS & -    & -    & 31.4 & 15.0 & 31.4 & 15.0 \\ \hline
    CVaR    & 56.3 & 1.65 & 44.8 & 1.65 & 49.8 & 3.65 \\
    CC      & 71.6 & 5.00 & 54.8 & 3.65 & 59.0 & 4.35 \\
    Prob.   & 93.6 & 5.60 & 68.0 & 4.50 & 72.7 & 5.60 \\
    \thickhline
    \multicolumn{7}{c}{14-generator Australian system} \\ \hline
    Det-UB  & 17.4 & 0.00 & 13.6 & 0.00 & 13.8 & 0.00 \\
    Det-RMS & -    & -    & 19.3 & 47.0 & 19.6 & 6.95 \\ \hline
    CVaR    & 41.7 & 2.35 & 30.6 & 0.65 & 32.9 & 1.00 \\
    CC      & 56.0 & 2.90 & 40.6 & 1.30 & 44.2 & 1.95 \\
    Prob.   & 79.1 & 3.10 & 60.3 & 2.05 & 61.2 & 2.85 \\
    \thickhline
    \end{tabular}
    \label{admlimit}
\end{threeparttable}
\vspace{-0.3cm}
\end{table}

For the deterministic criteria, the upper-bound formulation yields zero violations in both systems under the fixed critical-frequency MC tests. For the non-sinusoidal waveforms, the RMS formulation increases $\rho_{\max}$ by relaxing cross-harmonic peak alignment, but does not provide a strict peak-response boundary. This distinction is evident from the non-zero violation rates, particularly the 47.0\% observed for the square waveform in the Australian system. Hence, $\ccalM_{\mathrm{det}}^{\mathrm{rms}}$ provides a less-conservative surrogate for oscillation severity, while $\ccalM_{\mathrm{det}}^{\mathrm{ub}}$ is more appropriate when a strict peak bound is required.

Under frequency uncertainty, $\rho_{\max}$ increases substantially compared to the deterministic upper-bound limits. The CVaR consistently gives smaller limits than CC, reflecting its stronger treatment of upper-tail responses. The CC violation rates remain at or below the prescribed 5\% level, with more conservative outcomes in the Australian system due in part to the worst-case phase aggregation retained in the limit formulation. In contrast, the probabilistic criterion incorporates both frequency and phase uncertainty, resulting in the largest admissible magnitudes across all waveforms. Its empirical violation rates range from 4.5\% to 5.6\% in the IEEE 39-bus system and from 2.1\% to 3.1\% in the Australian system. The slight exceedance of 5\% may partly reflect the use of a common $\mathrm{CF}$. The resulting limits are approximately 3.1--3.4 and 4.4--4.5 times the deterministic UB limits in the respective systems, demonstrating the substantial influence of uncertainty assumptions on admissible fluctuation magnitudes.

To assess these limits in a system-level context, we compare the sum of admissible source magnitudes with each system's nominal capacity. For the IEEE 39-bus system, the deterministic UB limits correspond to approximately 1.4–1.9\% of the 6-GW capacity, increasing to 4.5–6.2\% under the probabilistic criterion. The corresponding ranges are only 0.4–0.6\% and 1.9–2.5\% in the 22-GW Australian system. Despite its larger capacity, the Australian system accommodates a smaller relative fluctuation level, indicating that admissible magnitudes depend strongly on modal characteristics and source-to-grid coupling rather than system size alone.

\begin{figure}[t]
    \centering
    \begin{tabular}{cc}
    \includegraphics[width=0.46\columnwidth]{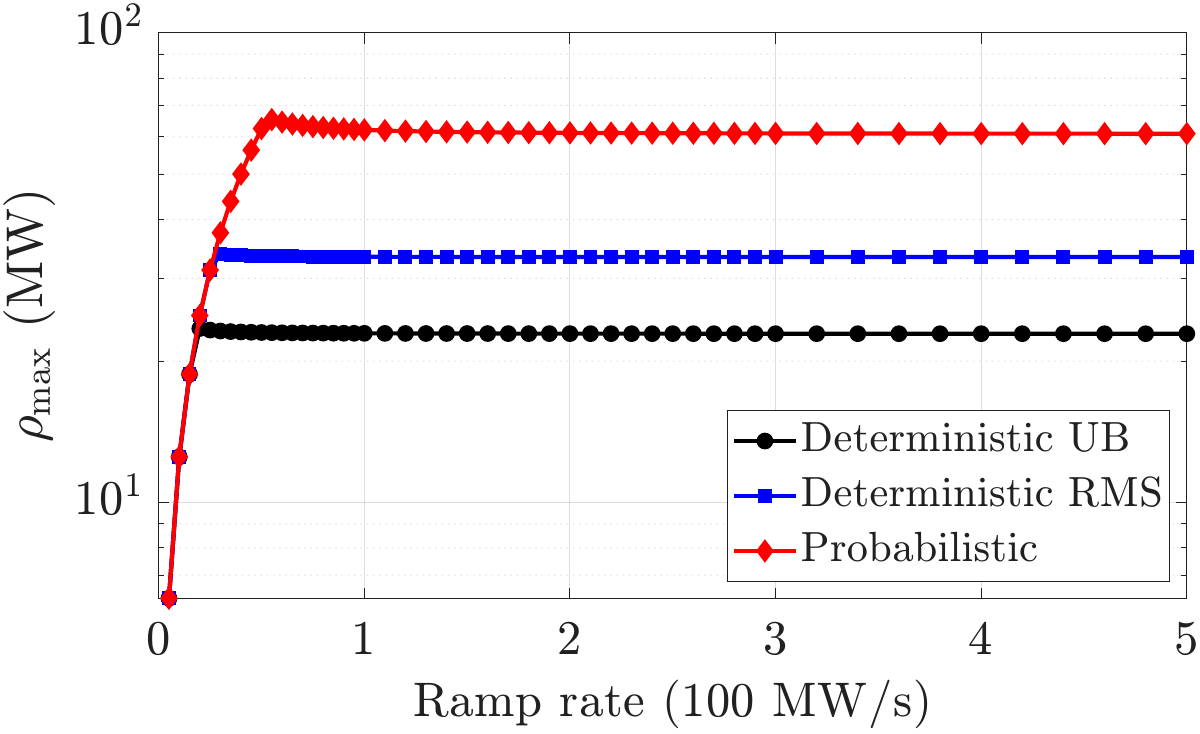} & \includegraphics[width=0.46\columnwidth]{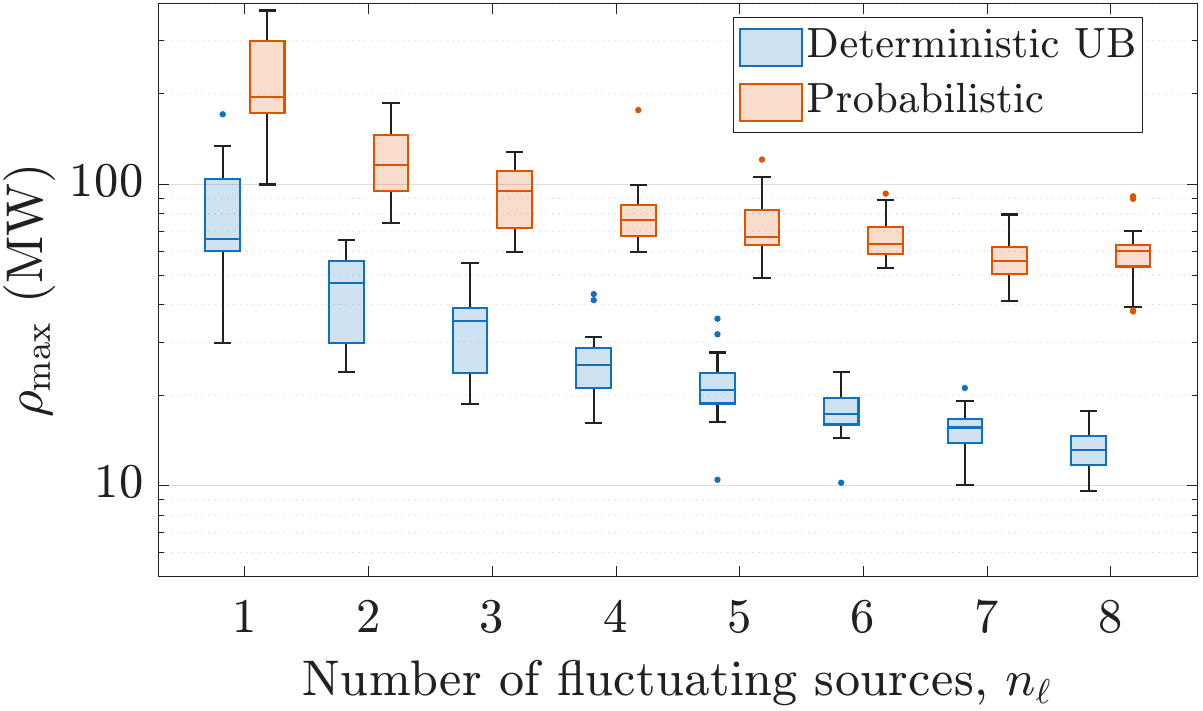}\\
    (a) & (b) \\
    \end{tabular}
    \caption{Impact of (a) ramp rates and (b) the number of forcing sources on admissible fluctuation magnitudes in the IEEE 39-bus system.}
    \label{rhomax}
    \vspace{-0.3cm}
\end{figure}

We further examine the effects of ramp rates and the number of forcing sources on $\rho_{\max}$ in the IEEE 39-bus system. Fig.~\ref{rhomax}(a) shows the impact of varying ramp rates on $\rho_{\max}$. At low ramp rates, $\rho_{\max}$ increases sharply before approaching a nearly constant level, with a minimal decrease at the highest rates. The initial increase is primarily driven by the waveform feasibility constraint, while the frequency-security constraint becomes dominant upon reaching a ramp-rate threshold. Together with Fig.~\ref{trapramp}, this result demonstrates that reducing ramp rates can decrease the oscillation severity while also restricting $\rho_{\max}$. Next, we vary the number of sinusoidal forcing sources from one to eight. For each of the 20 trials, the source locations are randomly selected from the load buses. Fig.~\ref{rhomax}(b) shows a substantial reduction in the admissible magnitude as the number of sources increases, with a faster decline at a smaller number of sources. The variation further indicates that the admissible limit depends on both the aggregated effects of forcing contributions and their spatial distribution across the grid.

\subsection{SC Optimization Results}
\begin{figure}[t]
    \centering
    \begin{tabular}{cc}
    \includegraphics[width=0.46\columnwidth]{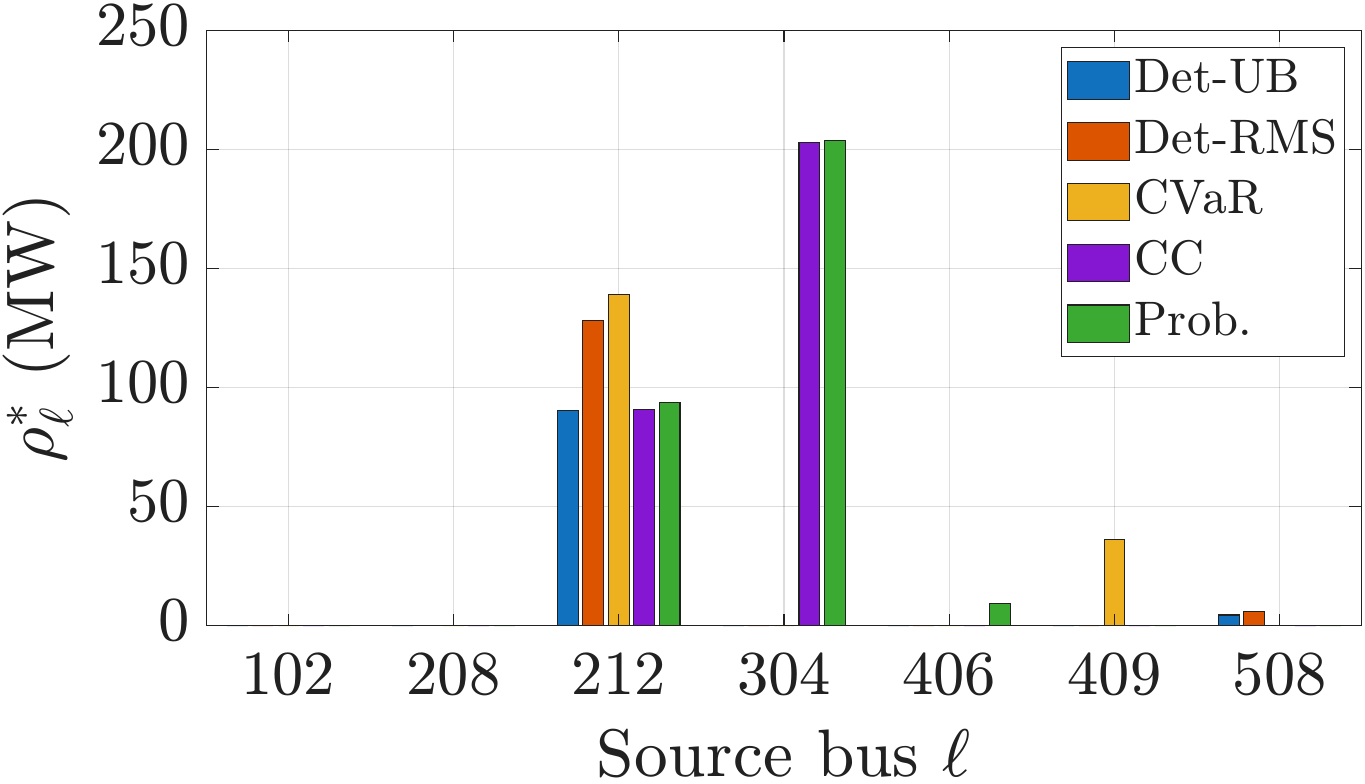} & \includegraphics[width=0.46\columnwidth]{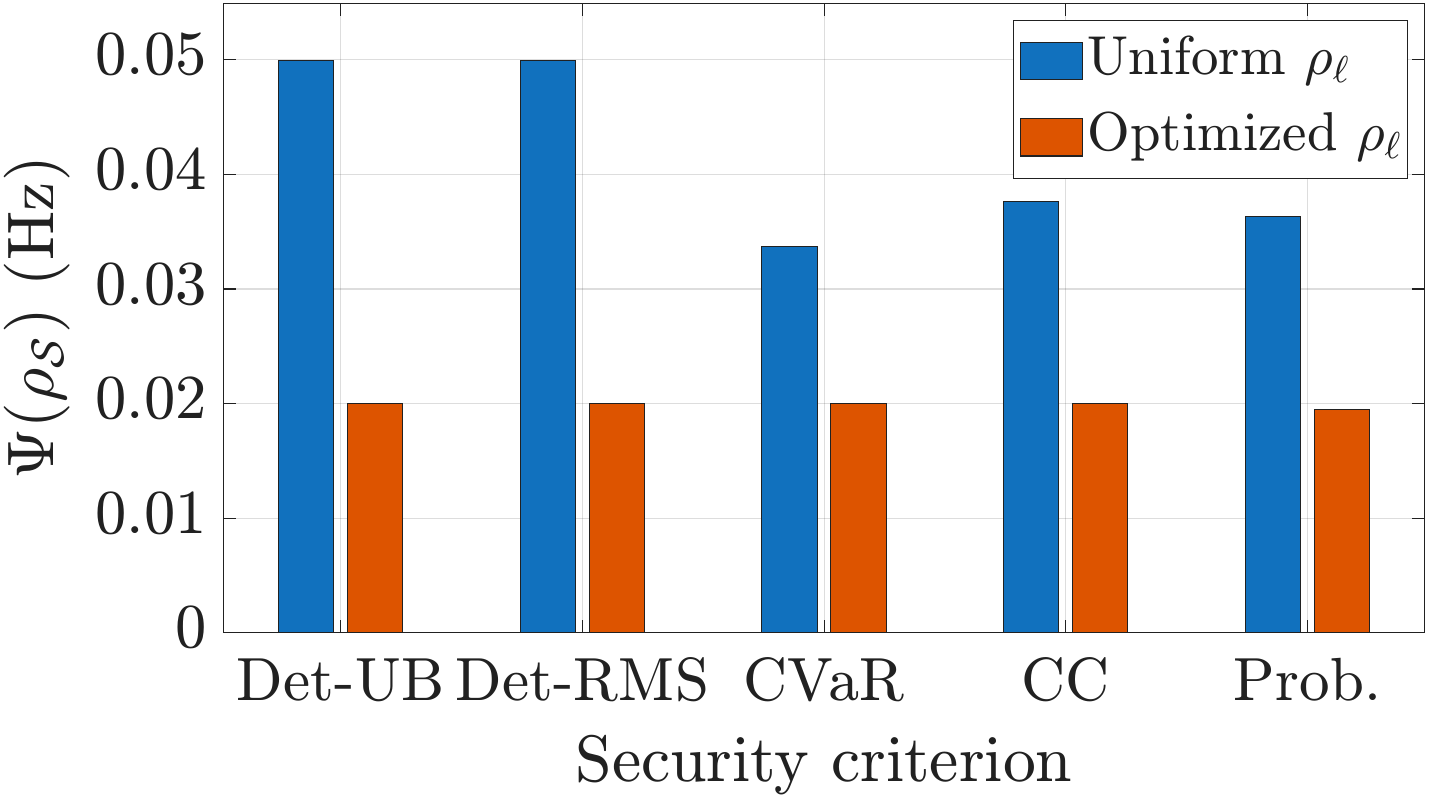}         \\
    (a)      &       (b) \\
    \multicolumn{2}{c}{\includegraphics[width=0.8\columnwidth]{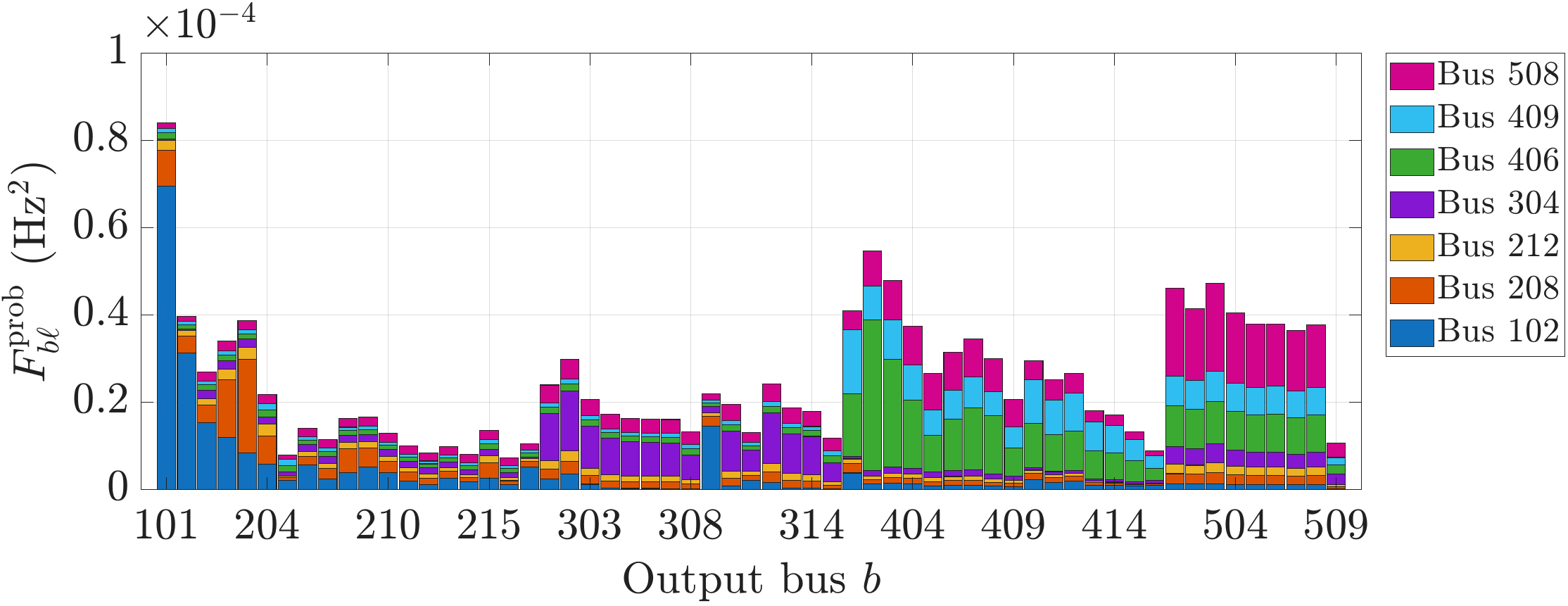}} \\
    \multicolumn{2}{c}{(c)}
    \end{tabular}
    \caption{SC optimization results in the Australian system: (a) optimized fluctuation magnitudes, (b) comparison of the security measure between uniform and optimized fluctuations, and (c) probabilistic source-to-output contributions under uniform source magnitudes.}
    \label{alloc}
    \vspace{-0.3cm}
\end{figure}

We finally evaluate the proposed SC optimization in the Australian system to maximize $\sum_{\ell}\epsilon_{\ell}$ in Sec.~\ref{sec5} for the trapezoidal waveform only, due to its higher complexity. Fig.~\ref{alloc}(a) compares the optimized fluctuation magnitudes at the seven source buses. The deterministic criteria favor Bus~212, with the RMS metric allowing a larger magnitude than the UB one. The CVaR result exhibits a similar preference while also assigning part of the fluctuation to Bus~409. In contrast, under the CC and probabilistic criteria, the dominant source shifts to Bus~304, with a smaller contribution from Bus~212. This difference highlights location-dependent grid sensitivities under each criterion: the deterministic constraints favor sources with smaller contributions based on the bus--frequency responses, while the frequency uncertainty affects the relative source sensitivities and can favor different locations. Fig.~\ref{alloc}(b) further compares the frequency-security measures between the uniform and SC-optimized source profiles, demonstrating the benefit of our grid-aware optimization accounting for the prescribed security requirement. The uniform cases produce $\Psi(\boldsymbol{\rho}_{\ccalS})$ values of approximately 0.034--0.050~Hz, while the SC-optimized ones remain close to the prescribed 0.02-Hz boundary. Reduced frequency severity can be achieved by  redistributing a total amount of fluctuation level to less sensitive locations.

To further interpret this spatial effect, Fig.~\ref{alloc}(c) shows the contribution of $F_{b\ell}^{\mathrm{prob}}$ in \eqref{fblprob} to $\ccalM_{\mathrm{prob}}$ under the uniform source profile. Across all input source locations, Buses~212 and 304 exhibit the smallest contributions, making the least impact on the maximum aggregate response variance \(J_b\) in \eqref{probenergy}. This is consistent with their larger fluctuation magnitudes allocated by the previous SC optimization, as these sources with smaller contributions can accommodate higher fluctuations without violating the system-wide frequency-security limit. We have also observed the same comparison on the $F_{b\ell}^{\rm det}$ values in \eqref{fbl} for the deterministic cases.

\section{Conclusion} \label{sec7}
This paper developed an analytical framework for assessing FOs induced by spatially distributed large-load fluctuations. We formulated deterministic and probabilistic severity metrics from the grid frequency response and used them to derive admissible fluctuation limits under different assumptions about frequency and phase uncertainty. The case studies showed that the resulting limits vary substantially with waveform characteristics, system dynamics, and the adopted risk treatment. The proposed SC optimization further showed that, for the same aggregate fluctuation level, redistributing source magnitudes according to their grid sensitivity can significantly reduce oscillation severity compared with uniform source magnitudes. These results highlight the importance of jointly considering fluctuation characteristics, uncertainty assumptions, and spatial source-to-grid coupling when establishing interconnection or operating limits for large fluctuating loads.

\ifCLASSOPTIONcaptionsoff
  \newpage
\fi
\bibliographystyle{IEEEtran}
\bibliography{bibliography}

@article{liu2024data,
  title={Data-driven forced oscillation localization using inferred impulse responses},
  author={Liu, Shaohui and Zhu, Hao and Kekatos, Vassilis},
  journal={Electric Power Systems Research},
  volume={234},
  pages={110759},
  year={2024},
  publisher={Elsevier}
}

@article{roh2007market,
  title={Market-based coordination of transmission and generation capacity planning},
  author={Roh, Jae Hyung and Shahidehpour, Mohammad and Fu, Yong},
  journal={IEEE Transactions on Power Systems},
  volume={22},
  number={4},
  pages={1406--1419},
  year={2007},
  publisher={IEEE}
}

@article{zhao2026dynamic,
  title={Dynamic Modeling of Data-Center Power Delivery for Power System Resonance Analysis},
  author={Zhao, Xingyu and Zhao, Junbo},
  journal={arXiv preprint arXiv:2604.06624},
  year={2026}
}

@article{charnes1959chance,
  title={Chance-constrained programming},
  author={Charnes, Abraham and Cooper, William W},
  journal={Management science},
  volume={6},
  number={1},
  pages={73--79},
  year={1959},
  publisher={INFORMS}
}

@article{rockafellar2002conditional,
  title={Conditional value-at-risk for general loss distributions},
  author={Rockafellar, R Tyrrell and Uryasev, Stanislav},
  journal={Journal of banking \& finance},
  volume={26},
  number={7},
  pages={1443--1471},
  year={2002},
  publisher={Elsevier}
}

@article{boyd1986multitone,
  title={Multitone signals with low crest factor},
  author={Boyd, Stephen},
  journal={IEEE transactions on circuits and systems},
  volume={33},
  number={10},
  pages={1018--1022},
  year={1986},
  publisher={IEEE}
}

@book{bonnans2006numerical,
  title={Numerical optimization: theoretical and practical aspects},
  author={Bonnans, J Fr{\'e}d{\'e}ric and Gilbert, J Charles and Lemar{\'e}chal, Claude and Sagastiz{\'a}bal, Claudia A},
  year={2006},
  publisher={Springer}
}

@article{marti2013multi,
  title={Multi-start methods for combinatorial optimization},
  author={Mart{\'\i}, Rafael and Resende, Mauricio GC and Ribeiro, Celso C},
  journal={European Journal of Operational Research},
  volume={226},
  number={1},
  pages={1--8},
  year={2013},
  publisher={Elsevier}
}

@article{spyrou2017benefits,
  title={What are the benefits of co-optimizing transmission and generation investment? {Eastern} interconnection case study},
  author={Spyrou, Evangelia and Ho, Jonathan L and Hobbs, Benjamin F and Johnson, Randell M and McCalley, James D},
  journal={IEEE Transactions on Power Systems},
  volume={32},
  number={6},
  pages={4265--4277},
  year={2017},
  publisher={IEEE}
}

@article{jensen2026easyrider,
  title={EasyRider: Mitigating Power Transients in Datacenter-Scale Training Workloads},
  author={Jensen, Dillon and others},
  journal={arXiv preprint arXiv:2604.15522},
  year={2026}
}

@article{al2020manipulation,
  title={Manipulation of static and dynamic data center power responses to support grid operations},
  author={Al Kez, Dlzar and Foley, Aoife M and Muyeen, SM and Morrow, D John},
  journal={IEEE Access},
  volume={8},
  pages={182078--182091},
  year={2020},
  publisher={IEEE}
}

@article{choukse2025power,
  title={Power stabilization for {AI} training datacenters},
  author={Choukse, Esha and others},
  journal={arXiv preprint arXiv:2508.14318},
  year={2025}
}

@article{hajiaghapour2024investigating,
  title={Investigating the cryptocurrency mining loads’ high penetration impact on electric power grid},
  author={Hajiaghapour-Moghimi, Mehran and others},
  journal={IEEE Access},
  volume={12},
  pages={153643--153663},
  year={2024},
  publisher={IEEE}
}

@techreport{follum2017power,
  title={Power system oscillatory behaviors: Sources, characteristics, \& analyses},
  author={Follum, James D and Tuffner, Francis K and Dosiek, Luke A and Pierre, John W},
  year={2017},
  number={1411936},
  institution={Pacific Northwest National Lab. (PNNL), Richland, WA (United States)}
}

@article{maslennikov2017dissipating,
  title={Dissipating energy flow method for locating the source of sustained oscillations},
  author={Maslennikov, Slava and Wang, Bin and Litvinov, Eugene},
  journal={International Journal of Electrical Power \& Energy Systems},
  volume={88},
  pages={55--62},
  year={2017},
  publisher={Elsevier}
}

@article{chevalier2018using,
  title={Using effective generator impedance for forced oscillation source location},
  author={Chevalier, Samuel C and Vorobev, Petr and Turitsyn, Konstantin},
  journal={IEEE Transactions on Power Systems},
  volume={33},
  number={6},
  pages={6264--6277},
  year={2018},
  publisher={IEEE}
}

@article{jiang2024forced,
  title={Forced oscillation source location in power systems using {MVMD}-assisted {DEF} in {TF} plane},
  author={Jiang, Tao and Liu, Bohan and Wang, Bin and Liu, Guodong and Li, Xue},
  journal={IEEE Transactions on Power Systems},
  volume={39},
  number={6},
  pages={6901--6913},
  year={2024},
  publisher={IEEE}
}

@techreport{nerc2021recommended,
  title={Recommended Oscillation Analysis for Monitoring and Mitigation Reference Document},
  author={{NERC Synchronized Measurement Working Group}},
  year={2021},
  institution={North American Electric Reliability Corporation (NERC), Atlanta, GA (United States)}
}

@techreport{nerc2010interconnection,
  title={Interconnection Oscillation Analysis},
  author={{NERC Synchronized Measurement Subcommittee}},
  year={2019},
  institution={North American Electric Reliability Corporation (NERC), Atlanta, GA (United States)}
}

@article{sun2022dynamic,
  title={Dynamic model and converter-based emulator of a data center power distribution system},
  author={Sun, Jingjing and Wang, Shuyao and Wang, Jingxin and Tolbert, Leon M},
  journal={IEEE Transactions on Power Electronics},
  volume={37},
  number={7},
  pages={8420--8432},
  year={2022},
  publisher={IEEE}
}

@article{jimenez2025data,
  title={Data center model for transient stability analysis of power systems},
  author={Jimenez-Ruiz, Alberto and Milano, Federico},
  journal={arXiv preprint arXiv:2505.16575},
  year={2025}
}

@techreport{ross2026electromagnetic,
  title={Electromagnetic Transient Modeling of Large Data Centers for Grid-Level Studies},
  author={Ross, Brett A and Follum, James D},
  year={2025},
  number={PNNL-38817},
  institution={Pacific Northwest National Laboratory (PNNL), Richland, WA (United States)}
}

@article{alshuaibi2022forced,
  title={Forced oscillation grid vulnerability analysis and mitigation using inverter-based resources: Texas grid case study},
  author={Alshuaibi, Khaled and others},
  journal={Energies},
  volume={15},
  number={8},
  pages={2819},
  year={2022},
  publisher={MDPI}
}

@article{valverde2025forced,
  title={Forced oscillations in power systems induced by {AI} workloads in data centers},
  author={Valverde, Gustavo and Pierrou, Georgia and Krajacic, Jovan and Hug, Gabriela},
  journal={Authorea Preprints},
  year={2025},
  publisher={Authorea}
}

@article{ko2026wide,
  title={Wide-area power system oscillations from large-scale {AI} workloads},
  author={Ko, Min-Seung and Zhu, Hao},
  journal={IEEE Transactions on Power Systems},
  year={2026},
  publisher={IEEE}
}

@inproceedings{biswas2025evaluating,
  title={Evaluating the risk to bulk power system reliability from large load induced oscillations},
  author={Biswas, Shuchismita and others},
  booktitle={2026 IEEE/PES Transmission and Distribution Conference and Exposition (T\&D)},
  pages={1--5},
  year={2026},
  organization={IEEE}
}

@misc{nerc2026assessment,
    author = {{NERC Large Loads Task Force}},
    title = {Assessment of Gaps in Existing Practices, Requirements, and Reliability Standards for Emerging Large Loads},
    url = {https://www.nerc.com/globalassets/our-work/guidelines/reliability/white-paper---assessment-of-gaps.pdf},
    institution = {North American Electric Reliability Corporation (NERC)},
    year = {2026}
}

@article{poolla2017optimal,
  title={Optimal placement of virtual inertia in power grids},
  author={Poolla, Bala Kameshwar and Bolognani, Saverio and D{\"o}rfler, Florian},
  journal={IEEE Transactions on Automatic Control},
  volume={62},
  number={12},
  pages={6209--6220},
  year={2017},
  publisher={IEEE}
}

@article{mishra2025inception,
  title={Understanding the inception of 14.7 {Hz} oscillations emerging from a data center},
  author={Mishra, Chetan and others},
  journal={Sustainable Energy, Grids and Networks},
  volume={43},
  pages={101735},
  year={2025},
  publisher={Elsevier},
  doi={10.1016/j.segan.2025.101735}
}

@article{hossain2026limiting,
  title={Limiting the Impact of {AI} Data Centers on Fatigue Life of Thermal Turbine Generators in the Grid: A Frequency-Domain Approach},
  author={Hossain, Fiaz and others},
  journal={arXiv preprint arXiv:2605.01173},
  year={2026}
}

@article{chaudhary2026predispatch,
  title={A Pre-Dispatch Resonance Safety Criterion for {AI} Training Clusters},
  author={Chaudhary, Chandan and Tiwari, Abanish and Pei, Yansong and Ben-Idris, Mohammed and Mitra, Joydeep},
  journal={arXiv preprint arXiv:2606.22096},
  year={2026}
}

@article{fan2026replicating,
  title={Replicating Real-World 23-{Hz} Oscillations Caused by Large Electronic Loads},
  author={Fan, Lingling and others},
  journal={arXiv preprint arXiv:2605.17190},
  year={2026}
}

@techreport{nerc2026risk,
  title={Reliability Guideline: Risk Mitigation for Emerging Large Loads},
  author={{NERC Large Loads Working Group}},
  institution={North American Electric Reliability Corporation (NERC)},
  year={2026}
}

@techreport{esig2026large,
    author = {Ahmed Rashwan and others},
    title = {Large Load Performance Requirements Current Practices and Recommendations},
    institution = {Energy Systems Integration Group},
    year = {2026}
}

@misc{caiso2026,
    author = {{California ISO}},
    title = {Large Load Considerations},
    url = {https://www.caiso.com/documents/presentation-large-load-considerations-stakeholder-information-session-feb-05-2026.pdf},
    year = {2026}
}

@techreport{nerc2025char,
    author = {{NERC Large Load Task Force}},
    title = {Characteristics and Risks of Emerging Large Loads},
    institution = {North American Electric Reliability Corporation (NERC)},
    year = {2025}
}

@article{chen2025electricity,
  title={Electricity demand and grid impacts of AI data centers: Challenges and prospects},
  author={Chen, Xin and others},
  journal={arXiv preprint arXiv:2509.07218},
  year={2025}
}

@misc{entsoe,
    author = {{ENTSOE}},
    title = {Network Code on Load-Frequency Control and Reserves},
    url = {https://eepublicdownloads.entsoe.eu/clean-documents/pre2015/resources/LCFR/130628-NC_LFCR-Issue1.pdf},
    year = {2013}
}

@misc{aemc,
    author = {{Reliability Panel AEMC}},
    title = {Frequency Operating Standard},
    url = {https://www.aemc.gov.au/sites/default/files/2024-01/Frequency%20Operating%20Standard.pdf},
    year = {2023}
}

\end{document}